\documentclass[conference]{IEEEtran}
\usepackage{algorithm}         
\usepackage{algorithmicx}     
\usepackage{algpseudocode} 
\usepackage{xspace}
\usepackage{threeparttable}
\usepackage{color}
\usepackage{wrapfig}
\usepackage{multirow}
\usepackage{listings}
\usepackage{comment}
\usepackage{booktabs}
\usepackage{makecell}
\usepackage{subcaption}
\usepackage{boxedminipage}
\usepackage{amsmath}
\usepackage{graphicx}
\usepackage{tabularx}
\usepackage{minibox}
\usepackage{pifont}
\usepackage{color}
\usepackage{xcolor}
\usepackage{subfiles}
\usepackage[utf8]{inputenc}
\usepackage[english]{babel}
\usepackage[cachedir=.]{minted}
\usepackage{xcolor}
\usepackage{ulem}
\usepackage{adjustbox}
\usepackage{xurl}
\usepackage{tcolorbox}
\newcommand{\bheading}[1]{{\vspace{1pt}\noindent{\textbf{#1}}}}
\newcommand{\ulitalic}[1]{{\vspace{1pt}\noindent\uline{\textit{#1}}}}
\newcolumntype{?}{!{\vrule width 1pt}}

\newcounter{note}[section]

\newcommand{\xmark}{\textcolor{red}{$\times$}}
\newcommand{\greencheck}{\textcolor{green}{$\checkmark$}}
\newcommand{\bluediamond}{\textcolor{blue}{$\diamond$}}
\colorlet{Mycolor1}{green!10!orange!90!}

\newcommand{\secref}[1]{\mbox{Sec.~\ref{#1}}\xspace}

\newcommand{\ignore}[1]{}

\newcommand{\etc}{\textit{etc.}\xspace}
\newcommand{\ie}{\textit{i.e.}\xspace}
\newcommand{\eg}{\textit{e.g.}\xspace}

\newcommand{\sysname}{\textsc{NetEcho}\xspace}

\newcounter{packednmbr}

\newenvironment{packeditemize}{
\begin{list}{$\bullet$}{
\setlength{\labelwidth}{0pt}
\setlength{\itemsep}{2pt}
\setlength{\leftmargin}{\labelwidth}
\addtolength{\leftmargin}{\labelsep}
\setlength{\parindent}{0pt}
\setlength{\listparindent}{\parindent}
\setlength{\parsep}{1pt}
\setlength{\topsep}{1pt}}}{\end{list}}

\newcounter{lessoncount}

\definecolor{greencell}{RGB}{146,210,14}
\definecolor{redcell}{RGB}{250,70,11}

\usepackage{pgfplots} 
\usepackage{tikz}
\pgfplotsset{compat=1.18}
\usepackage{array}
\usepackage{ragged2e}
\usepackage{pifont}
\usepackage{amssymb}
\usepackage{algpseudocodex}
\usepackage[table]{xcolor}
\usepackage{hyperref}
\usepackage{textcomp}

\newcolumntype{C}[1]{>{\Centering\arraybackslash}m{#1\textwidth}}

\renewcommand\arraystretch{0.9}
\newcommand{\cnum}[1]{%
  \tikz[baseline=(char.base)]{%
    \node[shape=circle,draw,inner sep=0.4pt,minimum size=0.8em,font=\scriptsize] (char) {#1};%
  }%
}
\newcommand{\circled}[1]{%
  \tikz[baseline=(char.base)]{%
    \node[shape=circle,draw,inner sep=0.4pt,minimum size=0.8em,font=\scriptsize] (char) {#1};%
  }%
}
\definecolor{highperf}{RGB}{200,245,200}    
\definecolor{medperf}{RGB}{255,255,224}     
\definecolor{lowperf}{RGB}{255,210,220}

\ifCLASSINFOpdf
\else
\fi
\begin{document}
%
\title{\sysname: From Real-World Streaming Side-Channels to Full LLM Conversation Recovery}


\author{
\IEEEauthorblockN{Zheng Zhang\IEEEauthorrefmark{1}\IEEEauthorrefmark{2}\thanks{Most of the work is done when Zheng Zhang was visiting SUSTech.},
Guanlong Wu\IEEEauthorrefmark{1},
Sen Deng\IEEEauthorrefmark{2},
Shuai Wang\IEEEauthorrefmark{2}\IEEEauthorrefmark{3}\thanks{\IEEEauthorrefmark{3}Corresponding authors.} and
Yinqian Zhang\IEEEauthorrefmark{1}\IEEEauthorrefmark{3}}
\IEEEauthorblockA{\IEEEauthorrefmark{1}Southern University of Science and Technology\\}
\IEEEauthorblockA{\IEEEauthorrefmark{2}The Hong Kong University of Science and Technology\\ 
Email:  zzhangla@cse.ust.hk, santiscowgl@gmail.com, \{sdengan, shuaiw\}@cse.ust.hk, yinqianz@acm.org}
}



%


\makeatletter
\def\footnoterule{\relax%
  \kern-5pt
  \hbox to \columnwidth{\hfill\vrule width 0.99\columnwidth height 0.6pt\hfill}
  \kern4.6pt}
\makeatother
\IEEEoverridecommandlockouts
\makeatletter\def\@IEEEpubidpullup{5.0\baselineskip}\makeatother
\maketitle

\begin{abstract}

In the rapidly expanding landscape of Large Language Model (LLM) applications, real-time output streaming has become the dominant interaction paradigm. While this enhances user experience, recent research reveals that it exposes a non-trivial attack surface through network side-channels. Adversaries can exploit patterns in encrypted traffic to infer sensitive information and reconstruct private conversations. In response, LLM providers and third-party services are deploying defenses such as traffic padding and obfuscation to mitigate these vulnerabilities.

This paper starts by presenting a systematic analysis of contemporary side-channel defenses in mainstream LLM applications, with a focus on services from vendors like OpenAI and DeepSeek. We identify and examine seven representative deployment scenarios, each incorporating active/passive mitigation techniques. Despite these enhanced security measures, our investigation uncovers significant residual information that remains vulnerable to leakage within the network traffic.

Building on this discovery, we introduce \sysname, a novel, LLM-based framework that comprehensively unleashes the network side-channel risks of today's LLM applications. \sysname is designed to recover entire conversations---including both user prompts and LLM responses---directly from encrypted network traffic. It features a deliberate design that ensures high-fidelity text recovery, transferability across different deployment scenarios, and moderate operational cost. In our evaluations on medical and legal applications built upon leading models like DeepSeek-v3 and GPT-4o,
\sysname can recover avg $\sim$70\% information of each conversation,
demonstrating a critical limitation in current defense mechanisms. We conclude by discussing the implications of our findings and proposing future directions for augmenting network traffic security.

\end{abstract}

\IEEEpeerreviewmaketitle



%

\section{Introduction}
\label{sec:introduction}

Large Language Models (LLMs) are increasingly deployed across diverse domains through two primary modes. 
First, major vendors offer services based on proprietary models, such as OpenAI's ChatGPT, DeepSeek's web interface~\cite{achiam2023gpt,liu2024deepseek}, and various API offerings. 
Second, developers build customized applications by curating domain-specific datasets~\cite{yang2024large,thirunavukarasu2023large} and integrating open-source foundation models via fine-tuning~\cite{lu2025fine} or RAG~\cite{lewis2020retrieval} frameworks. 
In both cases, these services deliver streaming, token-by-token responses.

Recent studies~\cite{weiss2024your,zheng2024inputsnatch,wu2025know} reveal that the token-by-token streaming of LLM responses leaks semantic information through network-level side channels. 
By analyzing encrypted traffic patterns---specifically packet timing and lengths---adversaries can infer parts of the model’s output or the user’s prompt (as detailed in \secref{subsec:existingattacks}).
To mitigate such side-channel risks, vendors have adopted two types of defenses.
First, active defenses~\cite{cloudflare2024mitigating} aim to fundamentally eliminate side-channel leakage by obfuscating traffic features through padding and batching.
Second, passive defenses reduce leakage granularity by aggregating multiple tokens into each packet instead of transmitting them one-by-one. While this approach does not fundamentally block the attack surface, it significantly increases the difficulty of conducting attacks~\cite{weiss2024your}.

However, in this paper, we show that the effectiveness of existing defenses remains limited in practice. 
We first conduct a real-world analysis of network traffic across a broad set of LLM applications. Despite that there exist varying deployment scenarios, where each scenario is incorporating defense methods, we observe a high volume of residual information, which may presumably leak private data.
On the other hand, we believe that existing attacks generally \textit{underexploit} the network traffic side channels. We show that by aggregating multi-dimensional network traffic and employing reasoning LLMs, we achieve highly effective attacks even under the state-of-the-art (SoTA) defenses.
We summarize key findings below:

\bheading{Real-World Study.} We surveyed SoTA defense mechanisms across different LLM application deployment scenarios. 
\begin{packeditemize}
\item  \textit{Active defenses are rare:} Only giant vendors implement padding~\cite{cloudflare2024mitigating}. Since active defenses require modifying streaming protocols and Server Sent Event(SSE) interfaces, most downstream applications cannot implement such measures.
\item \textit{Passive defenses remain weak:} Despite widespread API adoption, excessive tokens per chunk significantly degrade user experience~\cite{microsoft2024azure,chatbotui2024streaming}, limiting batching strength.
\item \textit{Frontend-backend configurations create diverse traffic patterns:} Different deployment setups exhibit significant variations in traffic patterns and residual information.
\end{packeditemize}

We summarize four kinds of inference backends and three mainstream frontends, and accordingly examine seven common deployment scenarios adopted by today's AI applications. We show that \textit{only one} of these scenarios are enforcing full-fledged defenses. 
To validate this observation, on medical~\cite{fansi2022ddxplus} and legal~\cite{tuggener2020ledgar} datasets, we collected both 10K data samples, and trained network trace-based classification models that can recover user conversation topics with accuracy of around 90\%. 

\bheading{Research Gaps.} 
There is an urgent need to comprehensively expose network side-channel risks of LLM applications, thereby promoting further defense efforts. Notice that existing attacks in this field are primarily effective in few scenarios~\cite{weiss2024your}, whereas other relevant attacks only leak basic information such as conversation topics~\cite{soleimani2025wiretapping,carlini2024remote,wei2024privacy}.  In short, we identify four research gaps in previous attacks:
\begin{packeditemize}
\item \textit{Isolated attack vectors}: Focus on specific mechanisms rather than comprehensive real-world scenarios.

\item \textit{Single traces}: Reliance on individual network trace analysis instead\! of\! a\! comprehensive,\! multi-dimensional\! trace\! viewpoint.

\item \textit{Partial recovery}: Limited to basic conversation information (topics or responses) rather than complete conversations.

\item \textit{Generation ability}:\! Fail to\! fully\! use SoTA\! LLMs\! to understand, reason,\! and generate high-quality textual conversations.
\end{packeditemize}

This paper introduces \sysname, a novel framework for recovering entire user-LLM conversations from encrypted network traffic. It uniquely combines two core techniques: trace-constrained text generation and iterative prompt refinement. Initially, \sysname uses a victim's network trace to retrieve similar (conversation, trace) samples from a pre-compiled database, adopting a RAG-like methodology. This database is created beforehand by the attacker, who acts as a regular user to probe the target application and log the conversation-trace pairs. These examples provide context for a local LLM to generate likely LLM responses constrained by the victim's traces. To further recover the user prompt, \sysname conducts a ``semantic gradient'' from the retrieved and generated content, which guides a reasoning model to iteratively rewrite and improve its guess of the user's original prompt. This iterative loop progressively refines the entire conversation, enabling high-fidelity recovery of both prompts and responses.

\sysname achieves highly accurate recovery on conversational contents, including both user prompts and LLM responses. Using local open-source models and remote reasoning models, our attack can be easily transferred to different scenarios by simply tweaking its configurations according to trace features. Across levels of characters, phrases, and semantics, we recovered 65\%/80\% of responses/prompts on medical data and 80\%/55\% on legal data, achieving up to 100\% attack success rate in certain scenarios.
Moreover, \sysname performs well on real-world data including imbalanced, large-scale scenarios and out-of-distribution samples, and demonstrates improvement in iterative refinement, highlighting the novelty and effectiveness.

\bheading{Contributions.} We make the following contributions:
\begin{packeditemize}
\item \textit{Systematic Analysis of Real-World Defenses}: We conduct the first comprehensive study of side-channel defenses across seven mainstream LLM deployment scenarios. We demonstrate that existing countermeasures are  ineffective in practice and fail to prevent information leakage from encrypted traffic.
\item \textit{A Novel Attack Framework}: We design and implement \sysname, a framework that pioneers a new attack paradigm. It for the first time integrates multi-dimensional trace analysis with LLM-driven reasoning, employing a RAG-like architecture and an iterative refinement loop to progressively reconstruct conversations with high fidelity.
\item \textit{First Demonstration of Full Conversation Recovery}: We elevate the threat model from partial data leakage to complete session recovery. \sysname is the first framework designed to recover both user prompts and LLM responses. Evaluations show it achieves an average attack success rate of 95\%, proving that current risks are critically underestimated.
\end{packeditemize}

\section{Background and Related Work} 
\label{sec:background}

\subsection{Construction of LLM Applications}
\label{subsec:llmapp}
With the growing capabilities of foundation models, an increasing number of domain-specific LLM applications are emerging~\cite{he2024foundation,henderson2023foundation}. Generally, LLM applications comprise two main components: backend inference engines and frontend user interface.
Due to intense competition in inference efficiency and accuracy, inference engines are limited to cloud providers' APIs (\eg, GPT, DeepSeek) and self-hosted frameworks (\eg, vLLM~\cite{kwon2023efficient}, SGLang~\cite{zheng2024sglang}). These interfaces largely follow the OpenAI standard~\cite{openai2024streaming}, creating a degree of uniformity in their operation. 

\begin{figure}[htbp]
    \centering
    \includegraphics[width=0.48\textwidth]{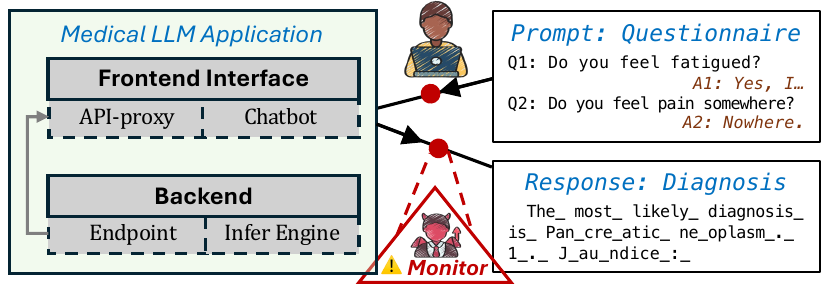}
    \caption{A medical LLM application which receives user prompts (questionnaire) and outputs preliminary diagnosis. Adversaries can monitor encrypted streaming.}
    \vspace{-0.1cm}
    \label{fig:layout}
\end{figure}

After generating results, the inference engine delivers results to users in two ways: (1) directly transmitting raw API outputs via REST protocols, or (2) rendering content through a frontend for better usability and presentation. Given the huge size and the hype of scaling law~\cite{zhao2023survey} of modern LLMs, responses cannot be generated instantly. Therefore, \textit{streaming} the output token-by-token is essential for a responsive user experience---and it is this streaming process that yields rich side channels. 
Fig.~\ref{fig:layout}  illustrates this architecture using a medical application. As the backend streams a diagnosis to user, an adversary can monitor the encrypted traffic. By analyzing packet timing and size patterns,  they can infer sensitive conversation content.

\subsection{Network Side Channels}
\label{subsec:existingattacks}

\bheading{Traffic Analysis.} Analyzing encrypted traffic is a mature research~\cite{dyer2012peek}. By passively observing traffic, prior work has inferred which class of LLM service a user is accessing and even which specific closed-source model. Alhazbi et al.~\cite{alhazbi2025llms} shows that different models exhibit stable inter-token-time signatures and can be distinguished in a fully black-box setting using only temporal vectors. Nazar et al.~\cite{nazari2024llm} demonstrate that packet-size and timing sequences alone can reveal if a service employs inference accelerators (e.g. speculative decoding~\cite{leviathan2023fast}). Active probing can identify service attributes even more accurately, including system prompts~\cite{song2024early} or proprietary templates. Therefore, via network side-channels, adversaries can passively and actively gather rich background knowledge.

\bheading{Traces from Optimization Mechanisms.} Certain optimization mechanisms produce distinctive traces. Wei et al.~\cite{wei2024privacy} report that speculative decoding leaves an observable signature, and Carlini and Nasr~\cite{carlini2024remote} list similar optimizations. These mechanisms generate multiple token candidates in parallel, then validate and accept the first $n$ tokens in one pass. This produces a distinguishable streaming pattern: packets burst at regular intervals separated by brief silences.
Using such trace, Soleiman et al.~\cite{soleimani2025wiretapping} train classifiers, recovering $\sim$70\% conversation topics.
Recently, we noticed that DeepSeek has deployed multi-token prediction (MTP) in production~\cite{liu2024deepseek}, which potentially introduces a similar side-channel.

\bheading{Trace from Token Length.} When an LLM streams, each token (or token chunk) is wrapped into an independent TLS record, leaving fine-grained length traces at the network layer. For a text, the tokenized sequence's lengths [$\text{token}_1$\_len, $\dots$, $\text{token}_n$\_len] leak through the encrypted connection. After capturing the packet-length series, an attacker can attempt to infer [$\text{token}_1$, $\dots$, $\text{token}_n$] from the lengths~\cite{weiss2024your}. Some providers already implement defenses: Cloudflare pads all traffic, while OpenAI and others only pad web-chatbot packages. API paths remain unpadded since they dynamically pack multiple tokens per encrypted packet, making current attacks~\cite{weiss2024your} ineffective.

\section{Problem Formulation}

\subsection{Threat Model}
\bheading{Attacker's Goals}.
We consider a comprehensive threat model where the attacker aims to recover the following properties:

\ulitalic{Topic}.  
Topic refers to fine-grained content categories~\cite{wei2024privacy,weiss2024your} that characterize the conversation subject. 
The attacker's goal is to classify conversations into specific topic labels with high accuracy.
Unlike prior work~\cite{carlini2024remote,zhang2024time} that focuses on coarse-grained classification (\eg, English vs France, Math vs. General questions), we target fine-grained categorization revealing specific user interests (\eg, in legal documents, distinguishing a clause's topic is ``Severability'' or ``Organizations'').

\ulitalic{Response}.  
Response refers to the complete model-generated output text. 
The attacker's goal is to reconstruct response content with at least 50\% semantic similarity to the original~\cite{weiss2024your}.

\ulitalic{Prompt}.  
Prompt refers to the user-issued input, often containing private or sensitive intent. 
The attack goal is to reconstruct prompt content with at least 50\% semantic similarity to the original, consistent with the response recovery goal.

\bheading{Attacker's Capability}.\label{sec:attack_cap}
Following prior work~\cite{soleimani2025wiretapping}, we assume a network-level adversary passively observing encrypted connections between victims and LLM services. They can identify the specific LLM model being used like prior-work~\cite{alhazbi2025llms}. 
The attacker also has access to the same LLM application: acting as normal users, they send similar queries and log the resulting conversation-trace within their budget.
For example, a co-worker using the same LLM application can easily collect such datasets while being curious about others' conversations, driving them to deploy monitors on public networks.

\subsection{Motivation}

\begin{table}[h]
\centering
\caption{Comparison with pervious network side-channel attacks,\,\greencheck,\,\bluediamond,\,\xmark\  \!denote support, partially, and not support.}\label{tab:comparison}\vspace{-0.1cm}
\renewcommand{\arraystretch}{1.15} 
\begin{threeparttable}
\begin{tabular}{c|c|c|c|c|c|c}
\hline
\centering & Feature & \cite{weiss2024your} & \cite{wei2024privacy} & \cite{soleimani2025wiretapping} & \cite{carlini2024remote} & \sysname \\
\hline
\multirow{3}{*}{\rotatebox{90}{{\!\!\!\textbf{Analysis}}}} & Explore wide deployment & \bluediamond & \xmark & \xmark & \bluediamond & \greencheck \\
\cline{2-7}
& Focus\! trace\! from\! optimization & \xmark & \greencheck & \greencheck & \greencheck & \greencheck \\
\cline{2-7}
& Focus\! trace\! from\! token\! length & \greencheck & \xmark & \xmark & \xmark & \greencheck \\
\hline
\multirow{4}{*}{\rotatebox{90}{{\!\!\!\textbf{Attack}}}} & Use modern LLM to attack & \xmark & \xmark & \xmark & \xmark & \greencheck \\
\cline{2-7}
& Recover conversation topic & \xmark & \greencheck & \greencheck & \bluediamond & \greencheck \\
\cline{2-7}
& Recover LLM response & \greencheck & \xmark & \xmark & \xmark & \greencheck \\
\cline{2-7}
& Recover user prompt & \xmark & \xmark & \xmark & \xmark & \greencheck \\
\hline
\end{tabular}
\end{threeparttable}
\vspace{-0.2cm}
\end{table}

Table~\ref{tab:comparison} compares our work with previous research. Overall, two types of gaps inspired our work:

\begin{packeditemize}
\item \textbf{Underexplored Threat}.
Studies~\cite{carlini2024remote,weiss2024your} have partially explored the vulnerability among deployments, pointed out the target LLM serving. However, similar to other studies, their analysis relies on individual network traces, leaving much of the attack surface as the \textit{tip of the iceberg}.
\end{packeditemize}
\ulitalic{Our Solution: Comprehensive Analysis}. Generated from one victim conversation, we analyze all effective traces across as many contemporary LLM applications as possible. More specifically, we explore factors that influence various trace characteristics (Sec.~\ref{sec:streaming_model}) and quantify the capability of different traces through evaluation (Sec.~\ref{sec:trace_eval}), thereby uncovering the substantial risks that lie beneath the surface.

\begin{packeditemize}
\item \textbf{Underpowered Recovery}.
Current work leaves significant room for improvement in both scope and methodology. Studies~\cite{soleimani2025wiretapping, carlini2024remote, wei2024privacy} can only recover conversation topics, substantially underestimating network side-channel risks. Some research~\cite{weiss2024your} ignores the semantic relationship between response and prompt, instead training traditional models for static recovery. This limits attack adaptability and fails to highlight the full impact such attacks can achieve.
\end{packeditemize}
\ulitalic{Our Solution: Generalizable Framework}. Given all the complex traces under different scenarios, we use one generic framework to fully recover the conversation. Specifically, by leveraging the response–prompt semantic relation, our approach iteratively refines the partially recovered results to gradually generate outputs with higher quality. This maximizes the practical impact of network side-channel attacks and, in turn, motivates the deployment of stronger defenses.

\begin{figure}[h]
    \centering
    \includegraphics[width=1\linewidth]{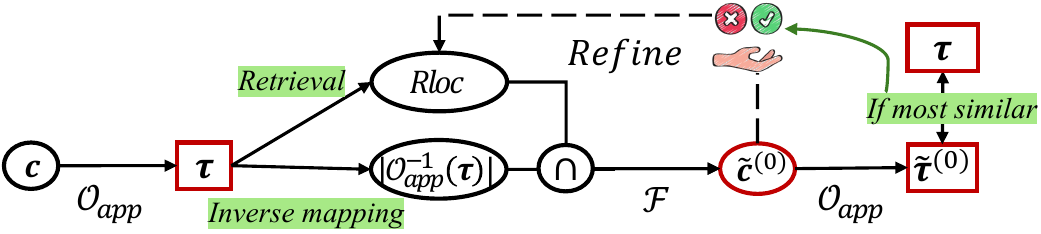}
    \caption{Given traces $\boldsymbol{\tau}$ generated from victim conversation $\mathbf{c}$, these traces enable the localization of the semantic space, in where LLMs can efficiently search for optimal recovery. Results serve as the refined reference of the next iteration, which further narrows down the semantic space.}\vspace{-0.1cm}
    \label{fig:problem}
\end{figure}

\subsection{Problem Formalization} 

\begin{table}[!htbp]
\centering
\caption{Notation table.}
\label{tab:notation}
\vspace{-0.1cm}
\renewcommand{\arraystretch}{1.25}
\resizebox{0.5\textwidth}{!}{%
\begin{tabular}{cl}
\hline
\textbf{Symbol} & \textbf{Description} \\
\hline
$c, \tau $ &  Victim conversation and corresponding trace \\
$\hat{c}, \hat{\tau}, \hat{c}^{(i)}, \hat{\tau}^{(i)}$ &  Probed conversation and trace for retrieval (at iteration $i$)\\
$\tilde{c}, \tilde{\tau},\tilde{c}^{(i)}, \tilde{\tau}^{(i)}$ & Recovered conversation and re-collected trace (at iteration $i$) \\
$Q, Q^{(i)}$ & Query database (at iteration $i$) \\
$\mathcal{O}_{\text{app}}, |\mathcal{O}_{\text{app}}^{-1}(\tau)|$ & Trace generation function / Search space of given trace \\
$\|\tilde{c}^{(i)}\|$ & Similarity with victim conversation: $\text{sim}(\tilde{c}^{(i)}, c)$ \\

\hline
\end{tabular}
}
\vspace{-0.2cm}
\end{table}

In this section, we formulate the adversary's core challenge: recovering the original conversation $\mathbf{c}$ solely from an observed network side-channel trace $\boldsymbol{\tau}$.
As shown in Notion Table~\ref{tab:notation}, let $\boldsymbol{\tau}$ be the network side-channel trace of a single conversation turn, where each $\tau_i$ encodes a packet’s length, arrival time, or other observable traffic features.  
Let $\mathbf{c}=\{\textsc{prompt},\textsc{resp}\}$ be a conversation with $m$-token prompt and $n$-token response.
Under a fixed tokenizer with $\mathbb{V}$ vocabularies, $\mathbf{c}$ resides in a discrete high-dimensional semantic space $\mathbb{C}_{\mathbb{V}}^{m+n}$.
The adversary can also query the LLM application to probe the semantic space. The corresponding queries and traces constitute the database $\mathbf{Q}=[\langle\hat{\mathbf{c}}_1,\hat{\boldsymbol{\tau}}_1\rangle,\dots,\langle\hat{\mathbf{c}}_b,\hat{\boldsymbol{\tau}}_b\rangle]$, where each $\hat{\mathbf{c}} = \{\hat{\textsc{prompt}}, \hat{\textsc{resp}}\}$.
The ultimate goal of the adversary is to find $\tilde{\mathbf{c}}$ that best approximates the true conversation, minimizing $\|\mathbf{c} - \tilde{\mathbf{c}}\|$.

\bheading{Trace Generation $\mathcal{O}_{app}$}. The network trace $\boldsymbol{\tau}$ is generated from the conversation $\mathbf{c}$ through a process we define as the function $\mathcal{O}_{app}$. This function, $\boldsymbol{\tau} = \mathcal{O}_{app}(\mathbf{c})$, models the entire pipeline of the LLM application, including its frontend and backend processing. This function is inherently a lossy projection because multiple distinct conversations ($\mathbf{c}_1 \neq \mathbf{c}_2$) can be mapped to the identical network trace $\boldsymbol{\tau}$.

The adversary's task is to solve the inverse problem: given a trace $\boldsymbol{\tau}$, find the original $\mathbf{c}$. Due to the lossy nature of $\mathcal{O}_{app}$, a single trace $\boldsymbol{\tau}$ corresponds to a set of potential original conversations. This set, denoted as $\mathcal{O}_{app}^{-1}(\boldsymbol{\tau})$, forms the adversary's search space. The difficulty of the attack is directly related to the size of this search space, $|\mathcal{O}_{app}^{-1}(\boldsymbol{\tau})|$. 

\bheading{Factors Affecting Attack Difficulty}. Under different interface $I \in \mathcal{I}$ and backend $B \in \mathcal{B}$ configurations of LLM applications, these mapping functions vary significantly, with each configuration corresponding to different search complexities. The introduction of defense mechanisms $D \in \mathcal{D}$ further expands the possible search space, extending the original mapping to a composite function $ D \circ \mathcal{O}_{app}$, causing the solution set of the inverse problem to grow substantially: $|D \circ  \mathcal{O}_{app }^{-1}(\boldsymbol{\tau})| \gg |\mathcal{O}_{app}^{-1}(\boldsymbol{\tau})|$, significantly increasing the search complexity to $\Omega(|D \circ  \mathcal{O}_{app }^{-1}(\boldsymbol{\tau})|)$. Since numerous LLM applications are closed-source, it is difficult to analyze and reverse-engineer the mapping functions from their fundamental implementations. One practical approach is to categorize scenarios and define scenario-specific mappings $\mathcal{O}_{app} = D \circ \mathcal{O}_{app}^{(I,B)}$, enabling targeted search space reduction strategies.

\bheading{Reducing Search Space via Semantic Seeker $\mathcal{F}$}.
To solve the inverse problem defined previously, we use LLMs as powerful semantic seekers. Our approach is built on a core hypothesis: trace similarity is indicative of semantic similarity. That is, if two conversations $\mathbf{c}_1$ and $\mathbf{c}_2$ are semantically close, their corresponding network traces $\boldsymbol{\tau}_1$ and $\boldsymbol{\tau}_2$ will also be similar $d(\boldsymbol{\tau}_1, \boldsymbol{\tau}_2) \propto d(\mathbf{c}_1, \mathbf{c}_2)$. Based on this, we reduce the vast search space in three sequential steps:

\ulitalic{Semantic Router via Classification}:
Real-world LLM applications typically involve conversations spanning dozens of topics, which naturally partition the semantic space into distinct clusters. Exploiting this structure, we develop a topic classification model that serves as a ``Semantic Router", enabling accurate retrieval of relevant samples from $\mathbf{Q}$ by narrowing down the search to the appropriate topic cluster.

\ulitalic{Semantic Filtering via Retrieval}: Then, we use the victim's trace $\boldsymbol{\tau}$ to find a small subset of reference examples $[\hat{\mathbf{c}}]$, from pre-probed queries database $\mathbf{Q}$ under the same topic. 
These examples are chosen because their traces are the most similar to $\boldsymbol{\tau}$. This step acts as a powerful filter, dramatically narrowing the search space from the entire universe $\mathbb{C}_{\mathbb{V}}^{m+n}$ to a much smaller, semantically relevant neighborhood, which we call $Rloc$. The reduction in complexity is significant, as $|Rloc| \ll |\mathbb{C}_{\mathbb{V}}^{m+n}|$.

\ulitalic{Trace-Constrained Generation}: After confining the search to the semantic neighborhood $Rloc$, we use the fine-grained features of the victim's trace $\boldsymbol{\tau}$ (\eg, token counts, character lengths) as hard constraints. An LLM is then tasked to generate text that is not only semantically consistent with the $\hat{\mathbf{c}}$ examples but also strictly adheres to the trace $\boldsymbol{\tau}$. This further prunes the candidate set, restricting the solution to the intersection of the semantic neighborhood and the set of conversations that could have produced the trace. Formally, the search complexity is $\Omega(Rloc\cap|\mathcal{O}_{app}^{-1}(\boldsymbol{\tau})|)$.

\bheading{Iterative Refinement for $\hat{\mathbf{c}}$}. 
The initial attack result, $\tilde{\mathbf{c}}^{(0)}$, provides a strong starting point, but its quality depends heavily on the initial reference set. We can further improve the attack's accuracy through an iterative feedback loop, which we term ``refinement'' in this paper. The core idea is to treat the output of each attack iteration as a new, high-quality data point to enrich our knowledge base. The process is as follows:

\ulitalic{Probe with New Candidate}: After generating serval candidates  $[\tilde{\mathbf{c}}^{(i)}]$ in iteration $i$, the adversary actively probes the target LLM application with it. This yields a new, re-collected trace $\tilde{\boldsymbol{\tau}}^{(i)} = \mathcal{O}_{app}(\tilde{\mathbf{c}}^{(i)})$ for each $\tilde{\mathbf{c}}^{(i)}$.

\ulitalic{Verify and Update}: The adversary then checks if this new trace $\tilde{\boldsymbol{\tau}}^{(i)}$ is a better match for the original victim trace $\boldsymbol{\tau}$ than the existing references. If this refinement condition is met (\ie, $d(\tilde{\boldsymbol{\tau}}^{(i)}, \boldsymbol{\tau})$ is smaller than the distances for previous references), the new pair $(\tilde{\mathbf{c}}^{(i)}, \tilde{\boldsymbol{\tau}}^{(i)})$ is added to the query database $\mathbf{Q}$, creating an updated, higher-quality database $\mathbf{Q}^{(i+1)}$.

\ulitalic{Narrowing the Search}: With an improved database, our employed retrieval model will produce a more accurate set of references $[\hat{\mathbf{c}}^{(i+1)}]$ in the next iteration. 
This leads to a more tightly focused semantic neighborhood $Rloc^{(i+1)}$. Consequently, the search complexity continues to decrease with each successful iteration $k$, as formalized by:
$\Omega(Rloc^{(k)} \cap |\mathcal{O}_{app}^{-1}(\boldsymbol{\tau})|) < \Omega(Rloc^{(0)} \cap |\mathcal{O}_{app}^{-1}(\boldsymbol{\tau})|)$.
This iterative process allows the attack to dynamically adapt and progressively converge on the true victim conversation.

\section{Analysis of Streaming Side-Channels}

In real-world deployments, LLM applications present complex network side-channel attack surfaces. Attackers face various side-channel traces due to different frontend-backend deployment configurations. 
In this section, we systematically summarize six attack surfaces upon seven  mainstream deployment scenarios and defense strategies, and quantify the capability of traces through topic recovery.

\subsection{Traces from Encrypted Traffic}
Encrypted traffic from a single response contains two hierarchical layers: valuable \emph{tokens} and observable \emph{packets}. Adversaries using packet capture tools can only observe the encrypted packets.  
On the backend, responses are typically generated token-by-token, with stable inter-token intervals~\cite{leviathan2023fast} due to substantial computational load, resulting in observable delays. However, the introduction of optimization mechanisms and batching defenses alter this pattern: at the packet layer, the observable units exhibit a \emph{token-group by token-group} pattern. In other words, regardless of how many tokens a packet contains, packets arriving at stable intervals within a single inference process can be logically grouped together. Moreover, by identifying the mapping between packet length and token length, an adversary can estimate the token length information contained within each packet. These characteristics can assist in inferring the content of streaming responses.

Overall, we categorize the information extracted from packets into two types of traces, namely \emph{Trace A} and \emph{Trace B}:
\begin{packeditemize}
    \item \textbf{Trace A: Token Count per Group.}  
    The adversary attempts to infer how many tokens arrive within the same interval at the packet layer. If one packet corresponds to a single token, tokens in packets arriving within a short interval can be considered as belonging to the same group. If a packet contains multiple tokens and packet arrival intervals remain stable, the adversary needs to estimate the number of tokens within each packet to obtain Trace~A.
    
    \item \textbf{Trace B: Character Length of Token(s).}  
    The adversary attempts to extract the character length of token(s) from packet length information. The granularity of Trace B also depends on the number of tokens per packet. If each packet corresponds to a single token, removing the fixed-length metadata enables the adversary to obtain the length of each token. If a packet contains multiple tokens, only the total character length of the token group can be determined.
\end{packeditemize}

\subsection{Factors Influencing Trace Characteristics}
\label{sec:streaming_model}

\bheading{Interface-Dependent Trace Characteristics.}
LLM applications primarily interact with users through two modalities: APIs and web chatbots. Although users perceive identical streaming token outputs, these two approaches exhibit distinct traffic patterns under adversarial monitoring.
We demonstrate that the token-to-packet mapping varies across interface types. Chatbots consistently exhibit one-to-one correspondence (single token per packet), whereas APIs tend to bundle multiple tokens per packets. This observation holds across OpenAI-style applications. Below, we take DeepSeek's official services \circled{A}Deepseek-Chatbot and \circled{B}Deepseek-API as examples to explain what information an adversary can obtain.

\begin{figure}[htbp]
    \centering
 
    \begin{subfigure}{0.85\linewidth}
        \centering
        \includegraphics[width=\linewidth]{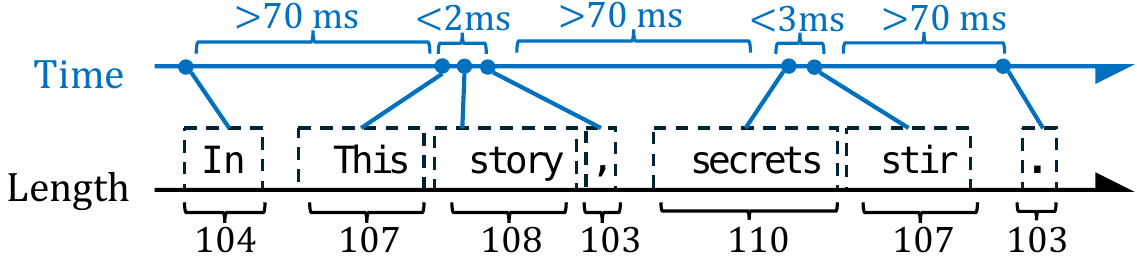}
        \caption{Packet characteristic of Deepseek Web-streaming.}
        \label{fig:package-web}
    \end{subfigure}
    \vspace{2mm}  
   \begin{subfigure}{0.85\linewidth}
        \centering
        \includegraphics[width=\linewidth]{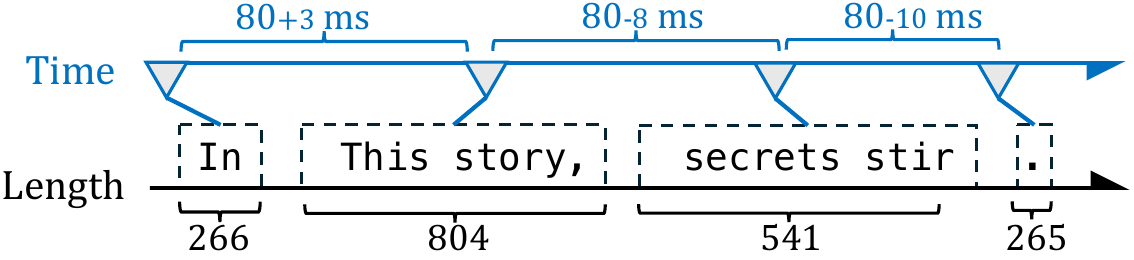}
        \caption{Packet characteristic of Deepseek API-streaming.}\vspace{-0.2cm}
        \label{fig:package-api}
    \end{subfigure}
    \caption{Deepseek traffic characteristics.}\vspace{-0.1cm}
    \label{fig:streaming_package}
\end{figure}

As shown in Fig.~\ref{fig:streaming_package}, DeepSeek-Chatbot exhibit traffic patterns composed of burst groups of packets arriving rapidly within short time intervals. In contrast, API connections display packets arriving at stable intervals with lengths approximating integer multiples of each other. These distinct interface characteristics enable different approaches for extracting Trace A and B. 
For the web-chatbot, Trace A is derived as [1,3,2,1]. Trace B is obtained by subtracting the fixed metadata length (\ie, 102 bytes) from each packet's total length. For API connections, token count per packet can be inferred from length multiples (\eg, 804 $\approx$ \emph{3} $\times$ 266), yielding the same Trace A vector [1,\emph{3},2,1]. Similarly, removing recurring metadata length (\eg, 264 bytes per token group) reveals the total character length per token group (\eg, $12 = 804 - 3 \times 264$).

\bheading{Backend-Induced Trace Difference}.
Backend configuration differences also create distinct variations. We use the proprietary application \circled{C}GPT-API as a study object to compare with \circled{B}Deepseek-API. While focusing on trace, we also analyze the related inference mechanisms in Appendix~\ref{app:infer_mechanism_spec}, whose findings are aligned with key findings of this section.

\ulitalic{Trace Reproducibility}.
In some backend models, certain traces fail to adequately reflect the  conversation information. We use trace reproduction to validate this observation. Specifically, we send the same query twice, repeat the process across multiple queries, and observe the resulting trace differences. For DeepSeek, under the same query, both traces exhibit only minor deltas, indicating strong reproducibility. However, for GPT, a small subset of queries exhibit bursts of packets arriving in rapidly, making Trace A unreproducible. In other cases, Trace B from the two queries is well-aligned, because the responses for the same prompt are highly similar.

\begin{figure}[htbp]
    \centering
    \begin{subfigure}{0.48\linewidth} 
        \centering
        \resizebox{0.85\linewidth}{!}{\includegraphics[trim=0 10pt 0 0, clip]{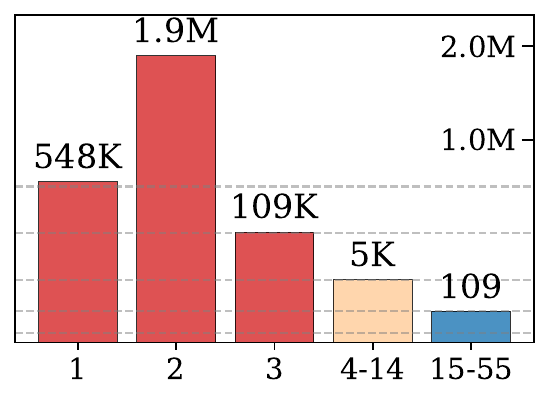}}
        \caption{Deepseek-v3-API.}
        \label{fig:ds-token-num}
    \end{subfigure}
    \begin{subfigure}{0.48\linewidth}
        \centering
        \resizebox{0.85\linewidth}{!}{\includegraphics[trim=0 10pt 0 0, clip]{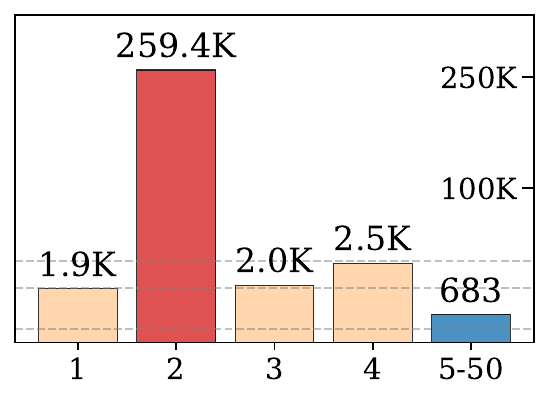}}
        \caption{GPT-4o-API.}
        \label{fig:gpt-token-num}
    \end{subfigure}
    \caption{Distribution of token count per packet.}\vspace{-0.2cm}
    \label{fig:token_distribution}
\end{figure}

\ulitalic{Trace Characteristic Variations}.
In the same type of trace, backend differences can also alter its characteristics. For Trace A, Fig.~\ref{fig:token_distribution} reveals distinct token-per-packet distributions between DeepSeek and GPT.
DeepSeek API's token counts mainly fall within [1,2,3], aligning with its multi-token prediction design.
In contrast, GPT API exhibits a divergence where 2-token packets constitute the overwhelming majority ($98\%$). 
The small fraction of outliers deviating from this pattern may be attributed to network noise, congestion, and jitter.
For Trace B, different tokenizers across foundation models lead to varying token segmentation lengths. For instance, the word [`severability'] is tokenized as [`sever', `ability'] by DeepSeek, while GPT produces [`se', `ver', `ability']. This difference manifests in Trace B as [5,7] v.s. [2,3,7]. 

\subsection{Attack Surfaces under Existing Defense }

\begin{figure}[htbp]
    \centering
    \includegraphics[width=1\linewidth]{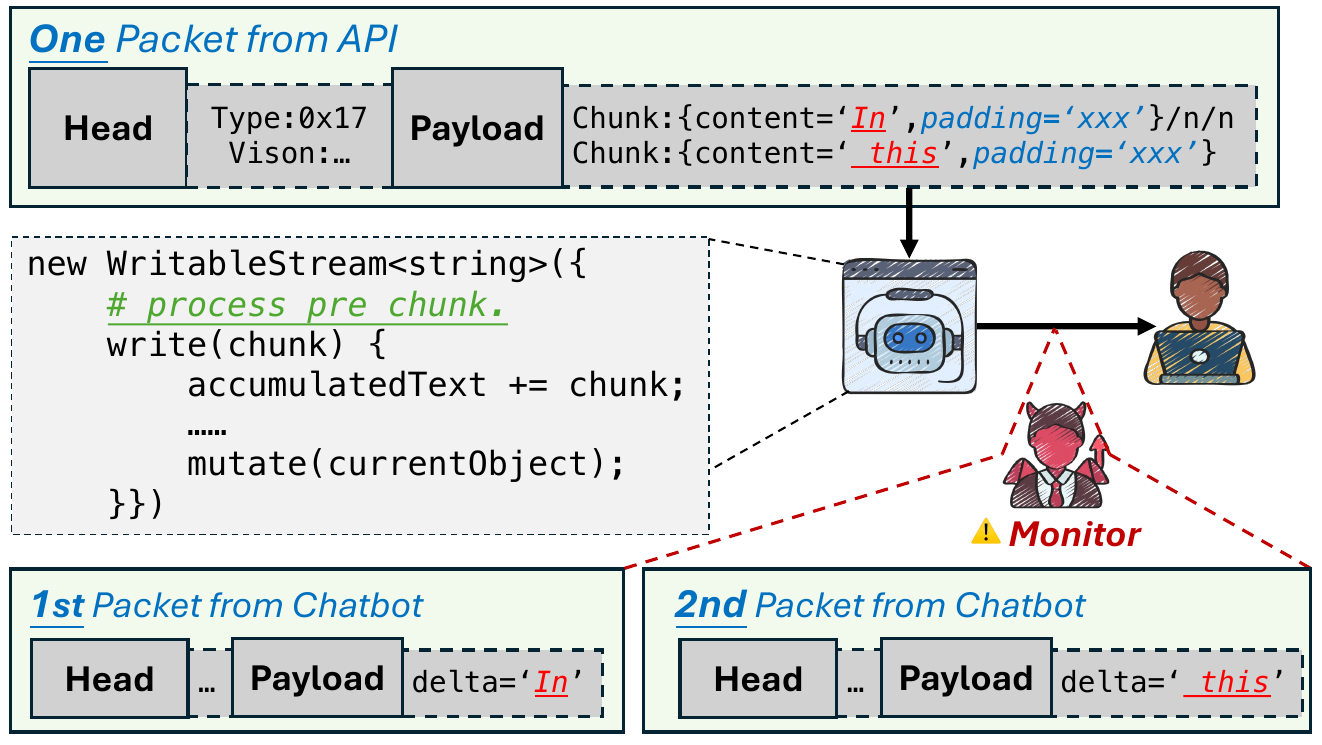}
    \caption{Even the upstream API adopts defense, the downstream interface~\cite{vercel2025ai}  still interacts with users token-by-token.}\vspace{-0.1cm}
    \label{fig:decode}
\end{figure}

To mitigate network side-channel attacks, vendors have implemented two categories of defensive countermeasures:

\bheading{Active defenses} refer to deliberately modifying the streaming protocol to obscure token-related features, typically by adding \emph{padding} to each packet and \emph{batching} multiple packets together. Fig.~\ref{fig:decode} illustrates an example of an LLM application whose upstream vendor applies active defenses. In this case, the vendor simply merges the packet information of two tokens and adds a random-length padding to each token.  
In real-world deployments, vendors such as OpenAI and Gemini have deployed active defenses for these \circled{D}industry giants' chatbot. For APIs, to the best of our knowledge, only Cloudflare explicitly claims to implement padding.

\textit{However, active defenses are costly and rarely adopted}.  
In web chatbots, introducing active defense requires customized modifications to the frontend's SSE protocol (\eg, buffering $n$ tokens, adding padding, modifying data structures, \etc).
This customization burden explains the limited adoption: popular frameworks like Next.js~\cite{vercel2025ai} and even LLM-specific ones like FastChat~\cite{zheng2023judging} lack such support, leaving most chatbots vulnerable.
Similarly, enabling active defense for APIs requires modifying the standard protocol to introduce padding fields. A non-standard variant of the OpenAI-style response may cause issues in production environments. The compatibility problems are often more challenging than the modification cost itself.

Fig.~\ref{fig:decode} illustrates a classic chatbot by combining an upstream vendor's API serving with open-source frontend frameworks. As shown, even if the upstream API applies padding or batching, an adversary on the user side can still obtain the finest-grained traces. Breaking down such scenarios by vendor's active defense type, we identify the chatbot using: \circled{E}GPT-based backends with batching, and \circled{F}Cloudflare-like backends with padding. In these cases, despite the upstream's active defenses, downstream applications remain unprotected.

\begin{table*}[t]
\centering\small
\setlength{\arraycolsep}{0.5pt} 
\renewcommand{\arraystretch}{0.2} 
\caption{Deployment scenarios of LLM applications.}
\vspace{-5pt}
\begin{threeparttable}
\resizebox{0.95\textwidth}{!}{%
\begin{tabular}{@{}%
C{.08} 
C{.1} 
C{.12} 
C{.17} 
C{.17} 
C{.05} 
C{.25}@{}} 
\toprule
\multirow{2}{*}{\textbf{Scenario}} & \multicolumn{2}{c}{\textbf{Upstream Backend}} & \multicolumn{2}{c}{\textbf{Downstream Applications}} & \multicolumn{2}{c}{\textbf{Existing Defense}} \\
\cmidrule(lr){2-3} \cmidrule(lr){4-5} \cmidrule(lr){6-7}
& \textbf{Infer Engine} & \textbf{Technique} & \textbf{Interface} & \textbf{Example} &\textbf{Type}  & \textbf{Describe} \\
\midrule
\circled{A} & \multirow{2}{*}[-0.5ex]{\parbox[c]{.1\textwidth}{\centering Open-source, {DeepSeek}/ LLaMA}} & \multirow{2}{*}[-1.2ex]{\parbox[c]{.1\textwidth}{\centering Multi-token prediction}} & Vendor-hosted Chatbot & DeepSeek/Silicon Flow's Chatbot & None & Minor Variant of the standard format \\
\cmidrule(l){1-1}\cmidrule(l){4-7}
\circled{B} & & & OpenAI-style API & DeepSeek/Silicon Flow's API~\cite{siliconflow2025} & Passive & Batching 1$\sim$3 tokens into one packet  \\
\cmidrule(l){1-7}
\circled{C} & \multirow{3}{*}[-2.7ex]{\parbox[c]{.1\textwidth}{\centering Proprietary, Gemini/\\Claude/{GPT}}} & \multirow{3}{*}[-2.7ex]{\parbox[c]{.1\textwidth}{\centering Cache, Buffering\tnote{a}}} & Vendor-hosted Chatbot & Gemini/Claude/ GPT's Chatbot & Active & Batching and padding to K bytes per packet \\
\cmidrule(l){1-1}\cmidrule(l){4-7}
\circled{D} & & & OpenAI-style API & Gemini/Claude/ GPT's API & Passive &Batching $\sim$2 tokens into one packet \\
\cmidrule(l){1-1}\cmidrule(l){4-7}
\circled{E} & & & Self-hosted Chatbot & Glass.health~\cite{glass2025health}  /Harvey AI~\cite{harvey2025} & Active & Upstream batching, No coverage for downstream apps \\
\cmidrule(l){1-7}
\circled{F} & Third-party LLM proxy & Padding & Self-hosted Chatbot & Cloudflare's LLM Application~\cite{cloudflare2025workersai} & Active & Upstream padding, No coverage for downstream apps \\
\midrule
\circled{G} & \multirow{2}{*}[-2ex]{\parbox[c]{.1\textwidth}{\centering {vLLM}/\\SGLang}} & Speculative decoding & OpenAI-style API & FastChat~\cite{zheng2023judging} & Passive & Dynamic tokens from speculative decoding \\
\cmidrule(l){1-1}\cmidrule(l){3-7}
\circled{F} & & Padding and Spec. Decoding & Self-hosted Chatbot & LMArena~\cite{lmsys2025arena}, Vercel AI~\cite{vercel2024playground} & Active &Upstream padding, No coverage for downstream apps \\
\bottomrule
\end{tabular}
}

\begin{tablenotes}
\footnotesize
\item[a] Speculated. Analysis in Appendix~\ref{app:infer_mechanism_spec}.
\vspace{-0.3cm}
\end{tablenotes}
\end{threeparttable}
\label{tab:deployment_defense_scenarios}
\end{table*}

\bheading{Passive defenses} leverage inherent features of the OpenAI SDK, \ie sending multiple tokens within a single packet, thereby reducing information granularity and indirectly increasing attack difficulty. While the rationale behind this design is unclear, we speculate that it may be a simple countermeasure against attack~\cite{weiss2024your}. Compared to the fixed pairing mechanism revealed in 2023—where each packet contained exactly two tokens—OpenAI now adopts a more dynamic packet granularity strategy. This change invalidates the pairing assumption~\cite{weiss2024your} for scenario\circled{C} like GPT API. Since OpenAI effectively sets the standard for LLM application protocols, most LLM vendors adopt the OpenAI SDK to provide responses in the OpenAI format. As a result, this passive defense mechanism is widely present in other LLM APIs, including but not limited to DeepSeek API.

\textit{Passive defenses merely increase attack difficulty without providing substantial protection}. The OpenAI API SDK sends multiple tokens within one packet; this  essentially casts the trace granularity from ``token by token'' to ``token-group by token-group''. Typical defenses of such kind are scenarios~\circled{B}\circled{C}\circled{G} represented by Deepseek-API, GPT-API and vLLM-API. While this introduces great amount of information loss to attackers, the risk persists. Soon in this paper, we show that these insufficient defensive measures prove highly vulnerable.

\bheading{Systematical Analysis across Deployments}.
Further to the conceptual analysis of deployment factors and defense for traces, we conduct a comprehensive and empirical study in real-world LLM applications. Under four kinds of inference backends and three mainstream frontends across dozens of applications, we aim to reveal the attack surface as comprehensively as possible.
As illustrated in Table~\ref{tab:deployment_defense_scenarios}, we categorize existing LLM applications into seven typical scenarios based on deployment, inference engine, and interface format. As previously clarified, six scenarios (\circled{A},\circled{B},\circled{D}-\circled{G}) have distinct defensive characteristics and potential vulnerabilities.

While some vendor's chatbots in \circled{A} still lack defensive measures, most \circled{C} vendor-hosted chatbots implement comprehensive protection. For instance, GPT/Claude/Gemini chatbots  force adversaries to observe only several thousands-byte packets. In contrast, their API endpoints remain vulnerable: OpenAI-style APIs (\circled{B},\circled{D},\circled{G}) rely on batching, which is insufficient facing powerful attack.

As previously discussed, frontend frameworks cannot leverage upstream active defenses. This fundamental limitation affects scenarios \circled{E} and \circled{G}. Notably, some third-party vendors modify packet contents, and users may also alter content fields through optional parameters, though this does not affect trace extraction capabilities, as detailed in Appendix~\ref{app:variants}.

\subsection{Evaluating Resistance and Quantify Capability of Traces }
\label{sec:trace_eval}

Building upon the systematic analysis, we have identified six attack surfaces. Using topic classification tasks, we now empirically validate that adversaries can extract semantic information through traces, penetrating existing defenses across these scenarios. Concurrently, recovery accuracy serves as the quantitative metric to assess trace capability.
 
\bheading{Settings}. 
We simulate realistic attack workflows. To this end, we 
construct query prompts from diverse datasets, submit them to target LLM applications, and collect Trace A and B. The collected data are fed to a lightweight classification model to infer semantic content from network traffic patterns.

\ulitalic{Dataset Selection}. 
We selected medical~\cite{fansi2022ddxplus} and legal~\cite{tuggener2020ledgar} datasets. As a practical setup, we aim at saving the attack budget, drawing from established literature to construct a limited set of 10k prompts with corresponding 10 labels. We organized these structured data according to ChatCAD~\cite{wang2024interactive,zhao2024chatcad+} specifications, creating diagnostic queries for cloud-based LLM applications to determine which disease from a given list best matches the user's symptoms. Appendix~\ref{app:dataset} reports the details of converting raw data into structured prompts.

\ulitalic{Victim LLM Applications}. 
For each vulnerable scenario (\circled{A}, \circled{B}, \circled{D}–\circled{G}), we select a corresponding victim LLM application (\cnum{1}–\cnum{6}) for evaluation.
\cnum{1}\cnum{4} represent downstream chatbot deployments scenarios \circled{A}\circled{D}: DeepSeek's official Web Chatbot and a GPT-based chatbot builded on Next-js.
\cnum{2}\cnum{3} (scenarios \circled{B}\circled{E}) are real-world API services—DeepSeek-v3 and GPT-4o—demonstrating passive defenses from two leading vendors.
\cnum{5}\cnum{6} (scenarios \circled{F}\circled{G}) simulate active defense settings by implementing padding logic in vLLM (using Qwen3-32B) with speculative decoding n=6 and n=3, since few vendors support such defenses in practice.

\bheading{Trace Collection.\footnote{The collected data also serve as data sources for analysis in Sec.~\ref{sec:streaming_model}.}} We monitored connections between users and remote APIs, collecting $\sim$5k traces from GPT/vLLM, and $\sim$10k from DeepSeek. To form training, validation and test sets (8:1:1), we reconstruct trace information from API plaintext results, meanwhile using Wireshark to monitor specific IP packets and filter application data to obtain traces. We truncate all long packets to six tokens to reduce outlier impact.

\bheading{Trace Combo.} Trace B values are discretely distributed across a large range, making training difficult on small datasets. That explains why prior work~\cite{soleimani2025wiretapping} only uses the first 24 token lengths for classification. Therefore, we consider using 1) Trace A alone and 2) Trace A+B combo, using the number of tokens per group and the average character length per token respectively. Notice that, if the classification accuracy over Trace A+B is higher than that of using Trace A alone, trace B shall contain  orthogonal information to trace A.  After truncating the sequence length at the 95\% percentile, we feed these two types of sequences as inputs to the model below.

\bheading{Classification Model}. 
We train classification models using corresponding labels and traces, with trace data as input and classification results as output. Our architecture primarily employs a LSTM model~\cite{hochreiter1997long}.
Due to the limited training set the adversary can obtain, the training process may overfit. Thus, we use data augmentation methods~\cite{shorten2019survey} to enhance classification accuracy.
For the cost, we report that training on RTX 4090 required $\sim$600 seconds per experiment.

\begin{table*}[t]
\centering\footnotesize
\setlength{\arraycolsep}{0.3pt} 
\renewcommand{\arraystretch}{0.2} 
\caption{Trace characteristics and topic recovery accuracy result across LLM deployment scenarios. Values */* denote Top-1/Top-3 accuracy (\%), and we mark the highest recovery accuracy in \colorbox{highperf}{green}, medium in \colorbox{medperf}{yellow}, and lowest in \colorbox{lowperf}{red}.}
\vspace{-5pt}
\begin{tabular}{@{}%
   C{.04}  
   C{.15}  
   C{.15}  
   C{.12}  
   C{.14}  
   C{.07} 
   C{.07}  
   C{.07} 
   C{.07}@{}}
\toprule
\multirow{2}{*}{\centering\textbf{Scen.}}&
\multirow{2}{*}{\centering\textbf{Exp.}}&
\multirow{2}{*}[0.8ex]{\parbox{.15\textwidth}{\centering\textbf{Trace A\\Granularity}}} &
\multirow{2}{*}[0.8ex]{\parbox{.12\textwidth}{\centering\textbf{Trace B\\Granularity}}} & 
\multirow{2}{*}[0.8ex]{\parbox{.13\textwidth}{\centering\textbf{Trace Projection\\in Search Space}}} &
\multicolumn{2}{c}{\textbf{Results via Trace A}} &
\multicolumn{2}{c}{\textbf{Results\,Trace\,(A+)B}} \\
\cmidrule(lr){6-7} \cmidrule(lr){8-9}
& & & & & \textbf{Medical} & \textbf{Legal} & \textbf{Medical} & \textbf{Legal} \\
\midrule
\circled{A} & \raisebox{-0.8ex}{\cnum{1}DeepSeek-Chatbot} &
   Medium, from multi-token prediction &
   High, per-token length &
   $|\mathcal{O}_{app}^{-1}(\boldsymbol{\tau}_A\cap \boldsymbol{\tau}_B)|$ &
   \raisebox{-0.8ex}{\cellcolor{medperf}73.3/91.2} & \raisebox{-0.8ex}{\cellcolor{medperf}54.5/79.2} & \raisebox{-0.8ex}{\cellcolor{highperf}98.2/99.5} & \raisebox{-0.8ex}{\cellcolor{highperf}92.9/98.9} \\
\midrule
\circled{B} & \raisebox{-0.8ex}{\cnum{2}DeepSeek-API} &
   Medium, from multi-token prediction &
   Medium, sum token-group length &
   $ |\mathcal{O}_{app}^{-1}(\boldsymbol{\tau}_A^{3}\cap \boldsymbol{\tau}_B^{3})|$ &
   \raisebox{-0.8ex}{\cellcolor{medperf}72.6/90.9} & \raisebox{-0.8ex}{\cellcolor{medperf}53.8/80.3} & \raisebox{-0.8ex}{\cellcolor{highperf}93.7/98.8} & \raisebox{-0.8ex}{\cellcolor{highperf}88.7/98.0} \\
\midrule
\circled{D} & \raisebox{-0.8ex}{\cnum{3}GPT-4o-API} &
   Low, mechanism not observed &
   Medium, sum token-group length &
   $|\mathcal{O}_{app}^{-1}(\boldsymbol{\tau}_B^{2})|$ &
   \raisebox{-0.8ex}{\cellcolor{lowperf}11.0/26.4} & \raisebox{-0.8ex}{\cellcolor{lowperf}10.1/27.2} & \raisebox{-0.8ex}{\cellcolor{medperf}73.0/87.3} & \raisebox{-0.8ex}{\cellcolor{medperf}80.7/94.4} \\
\midrule
\raisebox{-0.7ex}{\circled{E}} & \raisebox{-0.8ex}{\cnum{4}GPT-based-Chatbot} &
   Low, mechanism not observed &
   High, per-token length &
   $|\mathcal{O}_{app}^{-1}(\boldsymbol{\tau}_B)|$ &
   \raisebox{-0.8ex}{\cellcolor{lowperf}11.2/26.1} & \raisebox{-0.8ex}{\cellcolor{lowperf}10.5/28.8} & \raisebox{-0.8ex}{\cellcolor{highperf}92.0/97.3} & \raisebox{-0.8ex}{\cellcolor{highperf}89.2/97.6} \\
\midrule
\raisebox{-0.7ex}{\circled{F}} & \raisebox{-0.8ex}{\cnum{5}vLLM-n=6-Chatbot} &
   High, from speculative decoding(n=6) &
   High, per-token length &
   $|\mathcal{O}_{app}^{-1}(\boldsymbol{\tau}_A^{6}\cap \boldsymbol{\tau}_B)|$ &
   \raisebox{-0.8ex}{\cellcolor{medperf}77.7/91.4} & \raisebox{-0.8ex}{\cellcolor{medperf}64.9/86.0} & 
   \raisebox{-0.8ex}{\cellcolor{highperf}91.6/98.4} & \raisebox{-0.8ex}{\cellcolor{highperf}89.6/95.5} \\
\midrule
\circled{G} & \raisebox{-0.8ex}{\cnum{6}vLLM-n=3-API} &
   Medium, from spec. decoding(n=3) &
   Medium, sum token-group length &
   $|\mathcal{O}_{app}^{-1}(\boldsymbol{\tau}_A^{3}\cap \boldsymbol{\tau}_B^{3})|$ &
   \raisebox{-0.8ex}{\cellcolor{medperf}76.5/91.2} & \raisebox{-0.8ex}{\cellcolor{medperf}60.9/85.4} & 
   \raisebox{-0.8ex}{\cellcolor{medperf}87.4/95.1} & \raisebox{-0.8ex}{\cellcolor{medperf}85.6/93.9} \\
\bottomrule
\end{tabular}
\label{tab:attack_scenarios}
\end{table*}

\subsection{Results and Discussions} 

\bheading{Limited Defense Effectiveness.}
Results shown in Table~\ref{tab:attack_scenarios} demonstrate that neither active nor passive defense mechanisms can completely eliminate network side-channel leakage in real-world LLM deployments. For Exp. under active defense with upstream batching~\cnum{4} or padding~\cnum{5}, both Trace A and B from  still reflect semantic information, revealing that downstream LLM applications leak exploitable information during streaming interactions. Similarly, for Exp. under passive defense,~\cnum{2}\,\cnum{3}\,\cnum{6} all yield traces with high classification capability, revealing the inadequacy of current defense.

While most configurations successfully leaked semantic information, GPT-4o's Trace A proved ineffective---conflicting with prior~\cite{soleimani2025wiretapping} results---suggesting that OpenAI's streaming interface either employs more low-level optimizations or deliberately restructures packets to obscure traffic.

\bheading{Trace Capability Comparison Across Scenarios.}
The capability of different trace types varies significantly across scenarios.
By comparing the theoretical trace granularity with the accuracy metrics from topic recovery, we can quantify the capability of each trace type.  
For Trace A, the highest granularity case (\cnum{5}) achieves better recovery accuracy than all other scenarios.  
In contrast, for Trace B, where only the total token length per group is available in~\cnum{2}\,\cnum{3}\,\cnum{6} under the same dataset and target LLM, the recovery accuracy of (A+)B is significantly lower than that of the corresponding \cnum{1}\,\cnum{4}\,\cnum{5}.

Formally, we can measure the attack difficulty of different scenarios through the size of trace projection in sematic space,
$|\mathcal{O}_{app}^{-1}(\boldsymbol{\tau}_B)|$, which depends entirely on the trace granularity.
For instance, compared to the scenario \cnum{4}, traces in scenario \cnum{3} manifest as coarser-grained $ \boldsymbol{\tau}_B^{2}$ due to deployment, with the corresponding conversation search space $|\mathcal{O}_{app}^{-1}(\boldsymbol{\tau}_B^{2})|$ exhibiting exponential growth. 
As the result, from \cnum{4} to \cnum{3}, top-1 recovery accuracy dropped from $>$90\% to $<$80\%.

\bheading{Feasibility and Challenges of Complete Attack.}
In our topic recovery evaluation, we employ a classification model to project trace features into a high-dimensional semantic space, enabling topic classification. Put differently, this process only allows us to estimate the approximate projection region of the trace in semantic space, denoted as $|\mathcal{O}_{app}^{-1}(\boldsymbol{\tau}_B)|$. Within this region, however, nearly infinite candidate conversations lie, with the victim conversation hidden among them.
If further narrowing our search within $|\mathcal{O}_{app}^{-1}(\boldsymbol{\tau}_B)|$, we can identify candidates semantically close to the victim conversation.

Across different scenarios, trace characteristics exhibit complex combinations, posing challenges for expanding the attack scope from classification to full-text recovery. Because trace granularity varies by scenario, the mapping between traces and tokens is inherently dynamic. As a result, static frameworks like~\cite{weiss2024your} training separate language models with fixed trace characteristics are not suitable. This observation motivates us to design a generic framework that, with only configuration changes, can recover full conversation in diverse scenarios.

\section{Towards Full Conversation Recovery}

Following our assessment in previous section, this section aims to 
design systematic attacks, by using LLM's capability to reason traffice trace and generate private information. This helps uncover the network side-channel risk surface in LLM applications,  thereby promoting the further implementation of defenses. Below, we first describe the base of our attack framework, \sysname, and then extend it to other scenarios.

\label{sec:attack}
\begin{figure*}[t]
    \centering
    \includegraphics[width=1\linewidth]{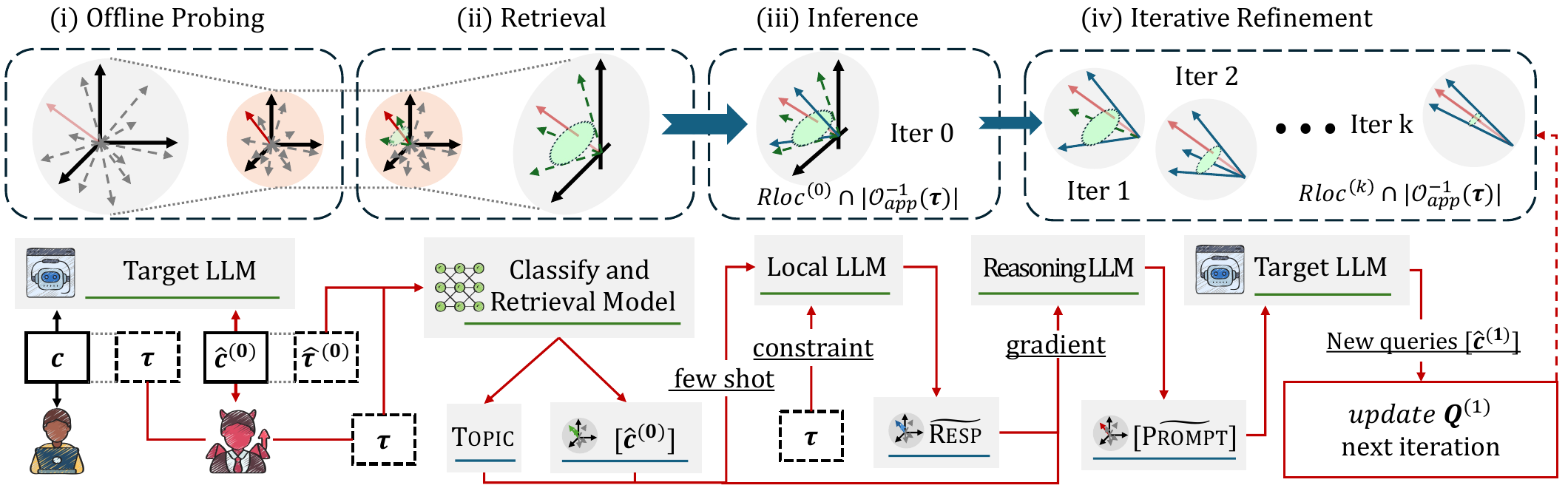}
    \caption{Framework of \sysname. It simultaneously depicts the step-by-step narrowing of the semantic space search scope during the attack process, and the interactions between  data and the LLM.}
    \label{fig:attack-framework}
\end{figure*}

\subsection{Base Framework}\label{sec:attack-framework}

Similarly to existing profiling-based side-channel attacks~\cite{yuan2022automated,kim2019make}, \sysname also consists of two phases: offline probing and online attacks. For clarity of presentation, we divide the online attack into three components: retrieval, inference, and interactive refinement.

\bheading{Offline Probe Stage ($\text{\romannumeral1}$)}. 
We maintain a design similar to related work~\cite{soleimani2025wiretapping}, namely conducting probes on LLM applications with specific datasets. For a given LLM application $\mathcal{O}_{app}$, the adversary prepares sufficient and uniformly distributed datasets, organizes them into prompts, and sends requests to the LLM application. From the LLM application's responses, the adversary can obtain both the plaintext query results and their corresponding traces $\langle\hat{\mathbf{c}},\hat{\boldsymbol{\tau}}\rangle$, 
where $\hat{\mathbf{c}}$ contains the prompt and response of this query.

Before launching the real attack, the adversary can make a limited number of such oracle calls and collect an auxiliary database $\mathbf{Q}$. 
The purpose of this step is to probe the entire semantic and trace spaces, providing sufficient training data for subsequent plaintext query retrieval based on trace similarity. As shown in Fig.~\ref{fig:attack-framework}($\text{\romannumeral1}$), gray represents the semantic space, orange represents the trace space, and there exists a certain mapping between them. Through sufficient probing, we can gain a good understanding of the distribution of these two spaces.
Consequently, the attacker holds the victim's trace $\boldsymbol{\tau}$, limited oracle access $\mathcal{O}_{app}$, and the a collection of queries $$\mathbf{Q}=[\langle\hat{\mathbf{c}}_1,\hat{\boldsymbol{\tau}}_1\rangle,\dots,\langle\hat{\mathbf{c}}_b,\hat{\boldsymbol{\tau}}_b\rangle] \in \langle\mathbb{C},\mathbb{T}\rangle.$$

\bheading{Retrieval ($\text{\romannumeral2}$)}.
We aim to establish the relationship between trace space $\mathbb{T}$ and semantic space $\mathbb{C}$ through a retrieval model $M_R$. The retrieval model $M_R$ takes the victim's trace $\boldsymbol{\tau}$ as input and returns several $\hat{\mathbf{c}}$ samples from query set $\mathbf{Q}$ that are closest to the victim's semantic information. That is, $$[\hat{\mathbf{c}}]=M_R(\boldsymbol{\tau},\mathbf{Q}).$$
Additionally, based on $M_R$, we can rank the similarity between each $\hat{\mathbf{c}}$ and $\mathbf{c}$ according to their traces. By feeding each $\hat{\mathbf{c}}$'s trace $\hat{\boldsymbol{\tau}}$ into $M_R$ for encoding and computing the similarity score with the victim's $\boldsymbol{\tau}$ encoding, we can obtain the semantic gradient by this score, formally defined as the function $\text{Rank}$:
$$
\textsc{gradv}=[\hat{\mathbf{c}}_{\mathrm{top}_1},\cdots,\hat{\mathbf{c}}_{\mathrm{top}_g}] = \text{Rank}(M_R,[\hat{\mathbf{\boldsymbol{\tau}}}]).
$$
Table~\ref{tab:retrieval_architecture} shows the structure of the retrieval model. Our core hypothesis is simple: distinct semantics generate unique network traffic patterns. To learn this pattern-to-semantic mapping, \sysname employs a dual-tower architecture, a design common in recommendation systems.

Specifically, the Trace Tower (using the classification model from Sec.~\ref{sec:trace_eval}) encodes traffic patterns, while the Text Tower (a MiniLM~\cite{wang2020minilm} Transformer) encodes the corresponding textual semantics. We train both towers jointly using a contrastive learning (InfoNCE) loss~\cite{oord2018representation}. This framework effectively teaches the Trace Tower to produce embeddings that match the Text Tower's embeddings for the same content, thereby establishing the crucial correspondence between traces and semantics. Through this trained model, $\sim$80\% of adversary's query traces can be successfully mapped to their semantic content.
Fig.~\ref{fig:attack-framework}($\text{\romannumeral2}$) illustrates this process. 
With only the victim's trace $\boldsymbol{\tau}$, the adversary uses the retrieval model to retrieve $\hat{\mathbf{c}}$ from $\mathbf{Q}$ that are closest to $\boldsymbol{\tau}$. 
This attempts to approximate the victim's conversation search space as $Rloc$, corresponding to the space enclosed by green arrows. The victim, represented by red arrows, lies within this space.

{\setlength{\tabcolsep}{1pt}
\begin{table}[htbp]
\centering\small
\caption{Retrieval dual-tower architecture.}
\begin{tabular}{@{}C{.01}C{.2}C{.2}C{.01}@{}}
\toprule
& \textbf{Text Tower} & \textbf{Trace Tower} &  \\ \midrule
 &Text Tnput      & Trace Tnput   &   \\ 
 &$\Downarrow$    & $\Downarrow$   &  \\ 
 &Transformer Encoder & Pretrained Classify Model & \\
& $\Downarrow$    & $\Downarrow$     &\\ \midrule
\multicolumn{4}{c}{\textbf{Contrastive Learning Head}} \\ 
\multicolumn{4}{c}{$L_2$ Normalize $\rightarrow$ Similarity Matrix $\rightarrow$ InfoNCE Loss} \\
\bottomrule
\end{tabular}
\label{tab:retrieval_architecture}
\end{table}}

\bheading{Inference ($\text{\romannumeral3}$)}. After determining the references, we narrow the semantics search space. We then seek to precisely recover the conversation content $\mathbf{c}$ from trace $\boldsymbol{\tau}$. The essence of this problem is to design a mapping such that the obtained $\tilde{\mathbf{c}}=\mathcal{F}(\boldsymbol{\tau},\mathbf{Q})$ approximates the real conversation $\mathbf{c}$ as closely as possible. 
We now show how to recover \textsc{resp} and \textsc{prompt} using two mappings, $\mathcal{F}_{1}$ and  $\mathcal{F}_{2}$, respectively.

\ulitalic{Recover Response}. After retrieval, the semantic bundle $Rloc$ still contains massive candidates.
To narrow the search, we need $\mathcal{F}_1$ to: (1) precisely understand semantic information to generate text within specified semantic ranges; (2) take traces as constraint and prune results that do not conform to the trace. We achieve these requirements by modifying locally deployed LLMs.

For challenge (1), we place the reference samples into organized prompt templates like \texttt{[System} \texttt{role}, \texttt{Reference} \texttt{samples}, \texttt{Confirmed} \texttt{label}, \texttt{Question]}. This way, the local LLM assistant has contextual knowledge, and the semantic information of output will also align with $\hat{\mathbf{c}}$.
For challenge (2), we force tokens at specified positions to have specified lengths during the generation process. We discuss the implementation in Sec.~\ref{sec:imple} in detail. Finally, given $\hat{\mathbf{c}}$ and trace $\boldsymbol{\tau}$, we can obtain the highest-confidence guess $\tilde{\textsc{resp}}$ for the response from the local LLM assistant:
$$\tilde{\textsc{resp}} = \mathcal{F}_1(\boldsymbol{\tau},[\hat{\mathbf{c}}]).$$

\ulitalic{Recover and Rewrite Prompt}. 
Prompts and traces are only connected through semantic relationships.  
We leverage the distance between the victim trace and $\hat{\boldsymbol{\tau}}$ to estimate the semantic similarity between the victim $\textsc{prompt}$ and $\tilde{\textsc{prompt}}$, \ie, $d(\mathbf{c}, \hat{\mathbf{c}} )\propto d(\boldsymbol{\tau}, \hat{\boldsymbol{\tau}})$.
Inspired by this, we adopt a gradient-based approach to guiding a reasoning model in recovering and rewriting prompts.  

Specifically, with the top-$g$ most similar $[\hat{\mathbf{c}}]$ samples from the query database, we can rank them by similarity, using function  $\textsc{gradv} =\text{Rank}(M_R, [\hat{\mathbf{c}}])$.  
The corresponding $\hat{\textsc{prompt}}$ from ${\mathrm{top}_g}$ to ${\mathrm{top}_1}$  form a semantic gradient whose direction points toward the victim \textsc{prompt}.  
We then encode this gradient into the prompt format \texttt{[System prompt, Example of Rewrite, Top-i Prompt, Target Response]} and feed it to a reasoning LLM. The reasoning LLM assistant with powerful infer capabilities will recognize this gradient and rewrite $g$ $\tilde{\textsc{prompt}}$ variants to further approximate the real victim prompt:
$$[\tilde{\textsc{prompt}}] = \mathcal{F}_2(\textsc{gradv},\tilde{\textsc{resp}}).$$

The method to determine which $\tilde{\textsc{prompt}}$ is closest to the victim prompt is straightforward. The adversary re-invokes $\mathcal{O}_{\text{app}}$ to obtain the corresponding $[\tilde{\boldsymbol{\tau}}]$ and $[\tilde{\mathbf{c}}]$, and selects the optimal $\tilde{\textsc{prompt}}$ based on the results of $\text{Rank}(M_R, [\boldsymbol{\tau}])$.

\bheading{Iterative Refinement ($\text{\romannumeral4}$)}.
Note that each recovery is essentially an LLM-based search for candidates near $\hat{\mathbf{c}}$ that satisfy the trace constraints. Since this search is centered around $\hat{\mathbf{c}}$, the key issue lies in continuously refining the quality of $\hat{\mathbf{c}}$, thereby progressively approximating the victim conversation. Meanwhile, we pursue minimal attack budget for the adversary, which means it is difficult to obtain high-quality reference samples in the initial attack phase. Fortunately, each recovered $[\tilde{\mathbf{c}}^{(i-1)}]$ could serve as the demand high-quality data in $i$-th iteration.

In the first iteration, we update $\mathbf{Q}^{(1)}$ with $[\tilde{\mathbf{c}}^{(0)}]$, and the retrieval results $[\hat{\mathbf{c}}] = M_R(\boldsymbol{\tau}, \mathbf{Q}^{(1)})$ are very likely to include $[\tilde{\mathbf{c}}^{(0)}]$. Building upon these improved ``anchors,'' $\mathcal{F}_{1}$ and $\mathcal{F}_{2}$ can generate higher-quality guesses $[\tilde{\mathbf{c}}^{(1)}]$. The effectiveness of this iterative refinement depends on the model capability, which we analyze in Sec.~\ref{sec:cross-tokenizer} and \ref{sec:iter_improve}.
Next, the adversary will repeat Steps $\text{\romannumeral1}$-$\text{\romannumeral3}$
until $\tilde{\mathbf{c}}^{(i)} \notin M_R(\boldsymbol{\tau},\mathbf{Q}^{(i +1)})$ or reach the pre-defined iteration threshold $k$. Fig.~\ref{fig:attack-framework}($\text{\romannumeral4}$) shows that the search space decreases to $R_{loc}^{(k)}\cap|\mathcal{O}_{app}^{-1}(\boldsymbol{\tau})|$ after $k$ iterations.

\subsection{Implementation in Real-World Settings}\label{sec:imple}
We present the basic attack pipeline in previous subsection. In reality, due to different trace characteristics and deployments, we present the following adaption to attack vectors. Note that these adaptions do \textit{not} affect the generalizability of the attack, as they do not involve model training processes like~\cite{weiss2024your}.

\bheading{Retrieval Vector}.
In the retrieval phase, we employ a dual-tower retrieval model $M_R$ to establish the mapping between trace space $\mathbb{T}$ and semantic space $\mathbb{C}$. 
The traces used by $M_R$ are completely adopted from the classification model employed in Sec.~\ref{sec:trace_eval}.
Given that trace effectiveness varies across scenarios, we employ Trace A+B for all experiments, as this combination performs the best.

\bheading{Local Model Assistant}. 
The local LLM assistant implements function $\mathcal{F}_1$, performing as a semantic seeker to efficiently find optimal recovery within the target semantic space. The local LLM must use a similar tokenizer to the victim model. Their token vocabulary spaces should largely overlap, allowing us to generate through a single forward inference.

Therefore, we select local models with tokenization algorithms and token vocabularies as similar with the target service as possible for each experiment. For Exp.\cnum{1}\cnum{2} with open-sourced DeepSeek v3, the local model is also the same. Due to the large size (671B) of the this model, we used DeepSeek-v2-chat-lite for simulation. 
For other experiments, especially for \cnum{3}\cnum{4}, we employed Qwen3-32B.
This is because attackers may have no information on close-model, and Qwen3 and GPT models are both developed based on extensions of the tiktoken tokenization framework~\cite{openai2023tiktoken}, leading to highly similar tokenization results for standard English text. We also analyze the cross-tokenizer robustness of our attack in Sec.~\ref{sec:cross-tokenizer}.

\bheading{Logits Constraint Setting}. 
To ensure that the responses generated by the local LLM assistant conform to trace constraints, we modify the logit matrix during the generation process. Based on different trace capabilities, we implement two constraint strategies: precise constraints for each token (Exp.\cnum{1}\cnum{2}\cnum{3}) and group constraints for token groups (Exp.\cnum{4}\cnum{5}\cnum{6}). For example, if trace A is [1,1] and trace B is [5,6], we process the logit matrix, keeping only tokens like [`\texttt{Hello}',`\_\texttt{world}'] with lengths 5 and 6. If trace A is [2] and trace B is [11], then we apply severe penalties to token group probabilities whose length sum does not equal 11 during the final token generation of this group.

\bheading{Remote Reasoning Model}. 
The remote reasoning model undertakes the role of function $\mathcal{F}_2$, responsible for rewriting prompts based on semantic gradient $\textsc{gradv}$. We deploy DeepSeek-R1 (Exp.\cnum{1}\cnum{4}\cnum{2}\cnum{6}) and GPT-o4-mini (Exp.\cnum{3}\cnum{5}) as two models with strong reasoning capabilities. This is to enable similar LLMs to attack themselves as much as possible. These models can recognize semantic gradients formed by similarity ranking and generate multiple candidate $\tilde{\textsc{prompt}}$ variants to approximate the real victim prompts. The generalization ability and creativity of the reasoning model directly determine the optimization effectiveness of the iterative refinement process.

\section{Evaluation}

In this section, we evaluate \sysname, addressing four main Research Questions (RQs):
\begin{tcolorbox}[
    colback=gray!10,
    colframe=black,
    boxrule=1pt,
    arc=3pt,
    left=2pt,
    right=2pt,
    top=2pt,
    bottom=2pt
]
\textbf{RQ1:} How effective is the attack? \textbf{RQ2:} 
How does it perform on real-world conditions, like scaling and OOD samples? \textbf{RQ3:} How effective is the adaptive improvement? \textbf{RQ4:} What is the resource consumption for attack?
\end{tcolorbox}

\begin{table*}[h]
\centering\small
\setlength{\arraycolsep}{2pt} 
\renewcommand{\arraystretch}{0.4}
\caption{Experimental results across different metrics (truncation at 300 tokens).}\label{tab:experimental_results}
\resizebox{1\textwidth}{!}{%
\begin{tabular}{@{}%
C{.02} 
C{.03} 
C{.06} 
C{.06} 
C{.06} 
C{.05} 
C{.05} 
C{.05} 
C{.05} 
C{.05} 
C{.05} 
C{.05} 
C{.05} 
C{.05} 
C{.05} 
C{.05} 
C{.05}@{}} 
\toprule
 & \multirow{2}{*}{\textbf{Goal}} & \multirow{2}{*}{\textbf{Metric}} & \multirow{2}{*}{\parbox[c]{.06\textwidth}{\centering \textbf{Local Output}}} & \multirow{2}{*}{\parbox[c]{.06\textwidth}{\centering \textbf{$\| \hat{\mathbf{c}}^{(0)}\|$}}} & \multicolumn{2}{c}{\textbf{Exp. \cnum{1}}} & \multicolumn{2}{c}{\textbf{Exp. \cnum{2}}} & \multicolumn{2}{c}{\textbf{Exp. \cnum{3}}} & \multicolumn{2}{c}{\textbf{Exp. \cnum{4}}} & \multicolumn{2}{c}{\textbf{Exp. \cnum{5}}} & \multicolumn{2}{c}{\textbf{Exp. \cnum{6}}} \\
\cmidrule(lr){6-7} \cmidrule(lr){8-9} \cmidrule(lr){10-11} \cmidrule(lr){12-13} \cmidrule(lr){14-15} \cmidrule(lr){16-17}
& & & & & $\|\tilde{\mathbf{c}}^{(3)}\|$ & \textbf{$\Delta$} & $\|\tilde{\mathbf{c}}^{(3)}\|$ & \textbf{$\Delta$} & $\|\tilde{\mathbf{c}}^{(3)}\|$ & \textbf{$\Delta$} & $\|\tilde{\mathbf{c}}^{(3)}\|$ & \textbf{$\Delta$} & $\|\tilde{\mathbf{c}}^{(3)}\|$ & \textbf{$\Delta$} & $\|\tilde{\mathbf{c}}^{(3)}\|$ & \textbf{$\Delta$} \\
\midrule
\multirow{13}{*}[-2ex]{\parbox[r]{.02\textwidth}{\centering \rotatebox{90}{Medicine}}} 
& \multirow{6}{*}[0.5ex]{\parbox[c]{.03\textwidth}{\centering \rotatebox{90}{\textsc{Resp}}}} 
& NED & 0.270 & 0.320 & 0.449 & +52\% & 0.376 & +27\% & 0.474 & +61\% & 0.845 & +186\% & 0.480 & +63\% & 0.326 & +10\% \\
\cmidrule(l){3-17}
& & ROUGE & 0.533 &  0.542 & 0.566 & +5\% & 0.460 & -14\% & 0.567 & +5\% & 0.822 & +53\% & 0.606 & +13\% & 0.472& -12\% \\
\cmidrule(l){3-17}
& & COS & 0.860 & 0.858 & 0.925 & +8\% & 0.890 & +4\% & 0.894 & +4\% & 0.955 & +11\% & 0.932 & +8\% & 0.882 & +3\% \\
\cmidrule(l){2-17}
& \multirow{6}{*}[1ex]{\parbox[c]{.03\textwidth}{\centering \rotatebox{90}{\textsc{Prompt}}}} 
& NED & - & 0.603 & 0.670 & +11\% & 0.641 & +6\% & 0.681 & +13\% & 0.703 & +17\% & 0.703 & +17\% & 0.654 & +8\% \\
\cmidrule(l){3-17}
& & ROUGE & - & 0.850 & 0.880 & +4\% & 0.850 & 0\% & 0.847 & 0\% & 0.882 & +4\% & 0.882 & +4\% & 0.847 & 0\% \\
\cmidrule(l){3-17}
& & COS & - & 0.943 & 0.961 & +2\% & 0.940 & 0\% & 0.937 & -1\% & 0.962 & +2\% & 0.962 & +2\% & 0.937 & -1\% \\
\cmidrule(l){2-17}
& \multicolumn{2}{l}{\textbf{ Success Rate}} & - & - & \multicolumn{2}{c}{100\%} & \multicolumn{2}{c}{93.3\%} & \multicolumn{2}{c}{95.6\%} & \multicolumn{2}{c}{100\%} & \multicolumn{2}{c}{97.3\%} & \multicolumn{2}{c}{78.6\%} \\

\midrule
\multirow{13}{*}[-5ex]{\parbox[c]{.02\textwidth}{\centering \rotatebox{90}{Legal}}} 
& \multirow{6}{*}[0.5ex]{\parbox[c]{.03\textwidth}{\centering \rotatebox{90}{\textsc{Resp}}}} 
& NED & 0.420 & 0.416 & 0.896 & +114\% & 0.707 & +69\% & 0.493 & +18\% & 0.857 & +105\% & 0.940 & +125\% & 0.719 &  +72\%\\
\cmidrule(l){3-17}
& & ROUGE & 0.577 & 0.625 & 0.878 & +46\% & 0.759 & +26\% & 0.621 & +3\% & 0.862 & +43\% & 0.943 & +57\% & 0.777 & +29\% \\
\cmidrule(l){3-17}
& & COS & 0.869 & 0.843 & 0.958 & +12\% & 0.937 & +9\% & 0.805 & -6\% & 0.894 & +4\% & 0.963 & +12\% & 0.857 & 0\% \\
\cmidrule(l){2-17}
& \multirow{6}{*}[1ex]{\parbox[c]{.03\textwidth}{\centering \rotatebox{90}{\textsc{Prompt}}}} 
    & NED & - & 0.312 & 0.451 & +45\% & 0.449 & +44\% & 0.442 &  +42\% & 0.449 & +44\% & 0.464 & +49\% & 0.466 &  +49\%\\
\cmidrule(l){3-17}
& & ROUGE & - & 0.471 & 0.573 & +22\% & 0.573 & +22\% & 0.589 &  +25\% & 0.605 & +28\% & 0.593 & +26\% & 0.583 & +24\% \\
\cmidrule(l){3-17}
& & COS & - & 0.470 & 0.547 & +16\% & 0.540 & +15\% & 0.562& +20\% & 0.580 &  +23\% & 0.571 & +21\% & 0.567 & +21\% \\
\cmidrule(l){2-17}
& \multicolumn{2}{l}{\textbf{ Success Rate}} &  &  & \multicolumn{2}{c}{100\%} & \multicolumn{2}{c}{90.0\%} & \multicolumn{2}{c}{93.5\%} & \multicolumn{2}{c}{100\%} & \multicolumn{2}{c}{100\%} & \multicolumn{2}{c}{95.1\%} \\

\midrule
\bottomrule
\end{tabular}
}\vspace{-1mm}
\end{table*}

\subsection{Evaluation Setup}

\bheading{Local Environment}. We deploy a dual L40 GPU machine as our local environment. This machine provides 96GB of GPU memory, enabling execution of models up to 32B parameters. Given hardware constraints, we focus on validating \sysname's attack effectiveness under fixed experimental settings rather than exploring various generation parameters like beams.
We employ Transformers~\cite{wolf-etal-2020-transformers} as local inference framework, which provides the \texttt{logits\_processor} functionality for flexible token-level generation control.

\bheading{Dataset Split and Others}.
For each experimental group, we randomly selected 100 data samples from the test set as victim conversations, then used the training set as retrieval queries. Other configurations, in particular victim LLM applications and dataset selection, are kept fully consistent with Sec.~\ref{sec:trace_eval}.

\bheading{Similarity Metrics}. We employ three complementary similarity metrics Normalized Edit Distance \textbf{(NED)}, ROUGE-1 F1 Score \textbf{(ROUGE)}, Cosine Similarity \textbf{(COS)} to comprehensively evaluate attack effectiveness, measuring text similarity at character, phrase, and semantic levels respectively. 
For convenience, we use the $\| \|$ notation to represent similarity with the victim conversation. For example, $\|\tilde{\textbf{c}}^{(3)}\|$ means the similarity between the final recovery and victim conversation.

\subsection{RQ1: Attack Effectiveness}

The topic recovery results have been presented in Sec.~\ref{sec:trace_eval}, demonstrating that traces in streaming transmission are sufficient for high-precision recovery of conversation topics. Here, we mainly focus on the recovery results for \textsc{prompt} and \textsc{resp}.

\bheading{Baseline}.
The recovery efficacy of \textsc{resp} is primarily determined by two factors: \textit{local LLM capabilities} and \textit{retrieved context quality} $||\hat{\mathbf{c}}^{(0)}||$. To evaluate the former, we directly pass the victim prompt to the local LLM and compute similarity scores between its \textit{direct output} and the victim response. To assess the latter, we measure the similarity between the responses in \textit{initially context} $\hat{\mathbf{c}}^{(0)}$
and the victim response. For \textsc{prompt} recovery, which exhibits greater independence, we utilize the similarity between the prompts in $\hat{\mathbf{c}}^{(0)}$ and the victim as the only baseline metric.

\bheading{Attack Parameters}.
We set the round threshold for iterative refinement to 3, with each rewriting operation generating 3 candidate $\hat{\textsc{prompt}}$s. During local LLM inference, we configure the beam search with a beam size of 30 and 5 beam groups, meaning we deploy 5 independent search beams, each maintaining 6 candidate sequences. We note that with more abundant computational resources, expanding the search space could significantly improve attack effectiveness, but this is not the core technical contribution of this work. Therefore, all subsequent experiments are conducted under identical parameter configurations to ensure result comparability.

\bheading{\textsc{Resp} Recovery}.
We present the experimental results in Table~\ref{tab:experimental_results}. Local Output represents the average similarity between local model outputs and victim outputs across experiments. There exists a significant discrepancy between the local LLM and target LLM. $\| \hat{\mathbf{c}}^{(0)}\|$ represents the quality of retrieval results. Compared to baseline, all experiments achieve an average 80\% improvement on the NED metric. This demonstrates that \sysname can effectively recover token-level information, approximating victim's actual textual outputs. At the semantic level, the recovery also achieves 12.5\% improvement.

\begin{figure}
    \centering
    \includegraphics[width=0.9\linewidth]{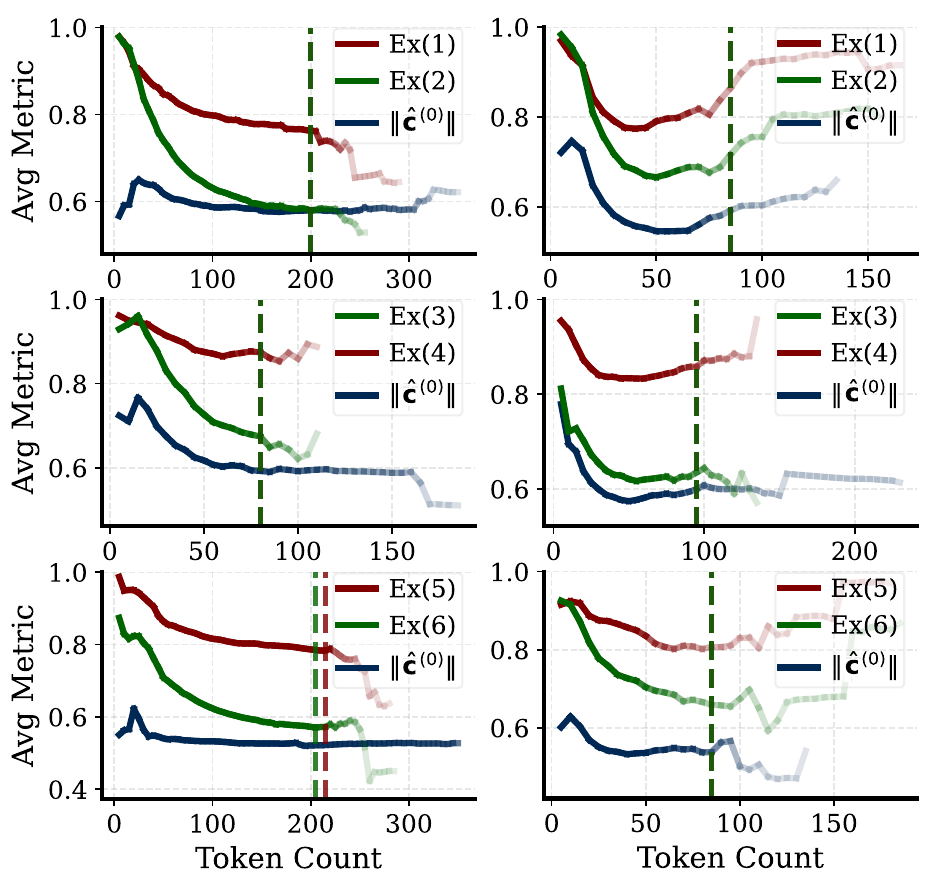}
    \caption{The recovery ratio $\|\tilde{\mathbf{c}}^{(3)}\|$ and baseline $\|\hat{\mathbf{c}}^{(0)}\|$ changes with token growth on medicine (left) and legal (right) datasets. Transparency represents the proportion of current data, and the dashed line indicates the 90\% percentile.}\vspace{-1mm}
    \label{fig:sim_by_token}
\end{figure}

\ulitalic{Drill-down Analysis.}
As shown in Fig.~\ref{fig:sim_by_token}, although 90\% of samples fall within 50\% of the maximum token length, longer texts still lead the LLM to generate more erroneous tokens that conform to the trace, resulting in performance decline. Nonetheless, most results remain better than the initial context $\|\hat{\mathbf{c}}^{(0)}\|$, confirming that \sysname effectively extracts information from traces despite the impact of error accumulation.

\begin{figure}[htbp]
    \centering
    \includegraphics[width=0.95\linewidth]{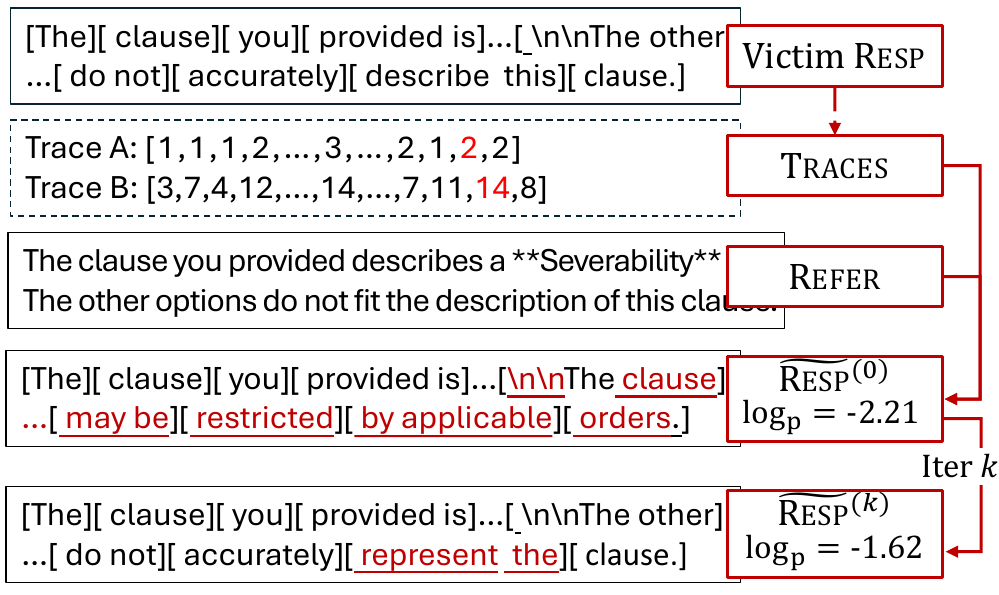}
    \caption{An example of \textsc{resp} recovery. Red indicates failed recovery; [token$_1$,$\dots$,token$_n$] denotes a group of tokens.}\vspace{-0.2cm}
    \label{fig:ex}
\end{figure}

\ulitalic{Case Study}.
Fig.~\ref{fig:ex} demonstrates an example of \textsc{resp} recovery. The trace effectively constrains the recovered $\hat{\textsc{resp}}$, ensuring that certain token groups (phrases within brackets []) match the character length of the victim \textsc{resp}. Meanwhile, $\hat{\mathbf{c}}$ provides semantic information to help the local LLM generate optimal solutions under these constraints. The log probabilities represent the quality of the constrained outputs, and the candidate with highest output quality serves as the final result. In Fig.~\ref{fig:ex}, since $e^{-1.62}\!>\!e^{-2.21}$, we choose $k$-th iteration's result.

\bheading{\textsc{Prompt} Recovery}.
\textsc{prompt} recovery presents greater challenges, given that its recovery mainly relies on context information. In the highly structured medicine dataset, since prompts in $\hat{\mathbf{c}}^{(0)}$ already closely match victim prompts, the improvement space is limited; \sysname achieves approximately 5\% performance improvement.
However, in the legal dataset, prompts manifest as independent legal clauses with relatively poor retrieval quality. Therefore, the recovered $\tilde{\textsc{resp}}$ plays a crucial role, enabling \sysname to additionally recover 28\% of \textsc{prompt} information.

\bheading{Attack Success Rate}.
We adopt the success criteria from~\cite{weiss2024your}, setting similarity 50\% as the threshold for successful attacks. After averaging the recovery results of \textsc{resp} and \textsc{prompt}, we calculate the proportion of successful attack trials, which shows that average 95\% of victim data are successfully recovered, with some experiments achieving 100\% success rate.

\bheading{Deployment Shift Robustness}.\label{sec:cross-tokenizer}
When attacking closed-source models, the local LLM faces two primary deployment shifts. 

\ulitalic{Output Style.} First, a significant gap in output quality between local/cloud LLMs~\cite{smith2025comprehensive}, which may be caused by differing system prompts or the use of post-training. This gap serves as our baseline in Table~\ref{tab:experimental_results}. For the same medical prompt, the NED similarity between local and cloud LLM response is only avg 0.27. Crucially, the attack improvement in this setting was not significantly lower than our identical-model upper bounds (Exp. \cnum{5}, \cnum{6}), showing \sysname's robustness to this shift.

\ulitalic{Mismatched Tokenizer.} To isolate the impact of this tokenizer shift from the output quality gap, we conducted a controlled experiment using two settings: (1) An upper-bound scenario (Exp. \cnum{5}, \cnum{6}) using the identical Qwen3-32B model as the target endpoint. (2) A realistic scenario employing DeepSeek-R1-Distill-Qwen-7B. This model, while comparable in capability to Qwen3-32B, uses a distinct variant of the Qwen3 tokenizer architecture with a smaller vocabulary (reduced from 11MB to 7MB).
By focusing exclusively on the initial attack (iteration $0$) performance on the legal dataset, we can isolate the tokenizer's effect. As shown in Table~\ref{tab:cross_tokenizer_effectiveness}, switching from an identical LLM to one with a highly similar tokenizer results in a negligible loss ($<$10\%) in attack accuracy. This demonstrates that \sysname is robust to tokenizer variations, a crucial factor for real-world deployment.


\begin{table}[htbp]
\centering\small
\setlength{\arraycolsep}{5pt}
\renewcommand{\arraystretch}{0.5}
\caption{Cross-tokenizer attack effectiveness comparison during initial attack period.}
\resizebox{0.9\columnwidth}{!}{
\begin{tabular}{@{}cccc||cccc@{}}
\toprule
& \multicolumn{3}{c||}{{Ds-R1-Distill-Qwen-7B}} & \multicolumn{3}{c}{{Qwen3-32B}} & \textbf{Avg $\Delta$} \\
\cmidrule(l){2-4} \cmidrule(l){5-7}
& {\textbf{NED}} & {\textbf{ROUGE}} & {\textbf{COS}} & {\textbf{NED}} & {\textbf{ROUGE}} & {\textbf{COS}} & \textbf{(\%)} \\
\midrule
\cnum{5} & 0.656 & 0.664 & 0.791 & 0.731 & 0.733 & 0.837 & 9.2\% \\
\cnum{6} & 0.433 & 0.535 & 0.767 & 0.452 & 0.545 & 0.790 & 3.1\% \\
\bottomrule
\end{tabular}
}
\label{tab:cross_tokenizer_effectiveness}
\end{table}

\subsection{RQ3: Real-World Scalability}
Real-world conversations rarely follow an ideal distribution. In practice, a single application can involve a massive number of labels, like numerous legal clause topic. From user's perspective, popular topics would be discussed more frequently, leading to highly imbalance of the sample distribution across topics.
Moreover, a significant amount of OOD data is invariably present within each topic. To evaluate NetEcho's performance under these extreme settings, we deep into Exp.~\cnum{3},\cnum{4}—the GPT-based legal scenario—to  expanded the number of conversational topics from 10 to 100 and probed an $80$k-sample imbalanced dataset to mimic a real-world application, which incurred an API cost of \$1,000.

\begin{figure}
    \centering
    \includegraphics[width=1\linewidth]{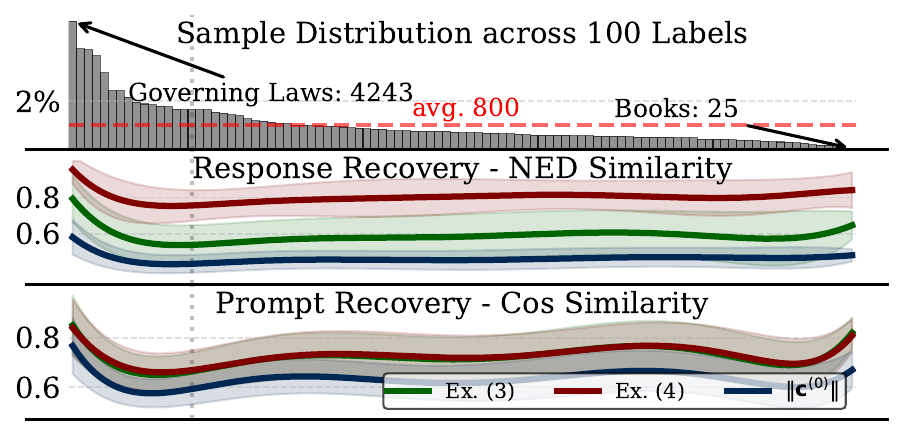}
    \caption{Recovery under imbalanced and large-scale dataset.}
    \label{fig:scaling}
\end{figure}

\bheading{Effectiveness of Large-Scaling and Imbalanced Data.} 
As shown in Fig.~\ref{fig:scaling}, the number of samples for the `Books' topic is merely 0.6\% of that for `Governing Laws'. This 100-imbalanced-labels scaling had an observable impact on recovery results. Following the same classification step as Sec.~\ref{sec:trace_eval}, we obtained a 92/98\% Top-1/Top-3 accuracy model in the web chatbot scenario and 80/93\% in the API scenario. 
We then selected 2\% samples from each label as the victim conversations. 
We focus on characteristic-level recovery of the response and semantic-level recovery of the prompt, as these metrics best reflect the recovery quality.
Fig.~\ref{fig:scaling} demonstrates that for the top 15\% of topics (which collectively account for 40\% of all samples), a clear, similar trend exists between the sample distribution and the recovery results. After that point, however, their trends diverge. The results from all experiments stop declining with the decreased budget, remain significantly better than the baseline and successfully recover 97.9\%/87.6\% victim conversations, confirming NetEcho's resilience under imbalanced scaling constraints.

\begin{figure}
    \centering
\includegraphics[width=1\linewidth]{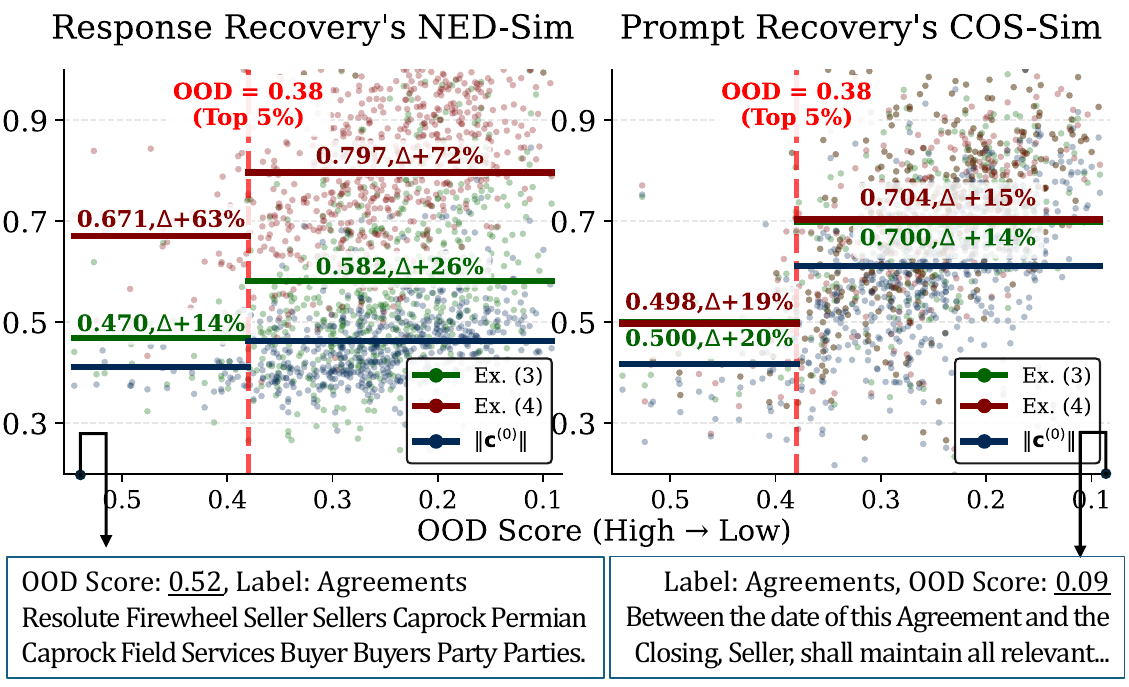}
    \caption{Performance under OOD samples.}
    \label{fig:ood}
\end{figure}

\bheading{Recovery of Out-of-Distribution (OOD) Samples.}
We scored victim samples for OOD score based on semantic distance to the nearest training sample, setting the 5\% boundary at 0.38. To prevent small-sample bias, we added the 2\% most OOD samples from each label.
Fig.~\ref{fig:ood} plots these results. Each point is an actual recovery, aggregated into averages for the top 5\% OOD samples and the remaining 95\%. 
As expected, OOD data recovery is lower than others, but the improvement over baseline is similar, especially for prompt recovery. The 5\% OOD samples still almost achieved no less than 50\% average recovery, showing \sysname generalizes well to extreme samples like the examples shown in the bottom of Fig.~\ref{fig:ood}.

\bheading{Cross-domain Scaling.}
The primary challenge in cross-domain scaling lies in the semantic router—that is, the classify step. Cross-domain scaling is similar to inner-domain scaling, though arguably more difficult because data distribution differences between domains are larger.
After routing, data entering the recovery process is located entirely within a single label's region in a single domain, consistent with the inner-domain setup. We trained a single classification model for the combination of legal and medical domains, which only introduced extra 0.5\% classification error. This demonstrates that \sysname has the potential to scale to multiple domains.

\subsection{RQ3: Adaptation Improvement}

\begin{figure}[htbp]
    \centering
    \includegraphics[width=0.7\linewidth]{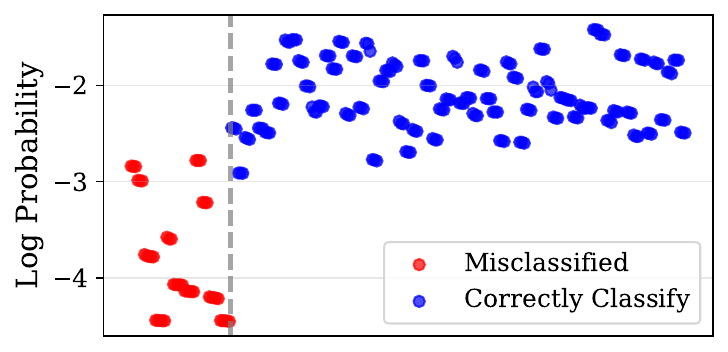}
    \caption{Log probabilities of generation for correctly classified samples (blue) and incorrectly classified samples (red).}\vspace{-2mm}
    \label{fig:logp}
\end{figure}

\bheading{Misclassification Correction.}
During \textsc{resp} recovery, the classification-based model $M_R$ may retrieve misclassified results whose semantics significantly differ with conversation $\mathbf{c}$.
A clear signal of misclassification is the notably lower confidence score output during trace-constrainted LLM generation, as shown in Fig.~\ref{fig:logp}.
P-test confirms that the log-probability distributions of correct and incorrect samples are statistically distinct. Accordingly, we treat samples with confidence below $\mu - \delta$ as erroneous and retrieve new reference based on the top-3 classification results.

\bheading{Iterative Improvement.}\label{sec:iter_improve}
In each refinement round, new information is input into the system. As shown in Table~\ref{tab:improvement_comparison}, compared to the initial recovery results $\tilde{\mathbf{c}}^{(0)}$, the final $\tilde{\mathbf{c}}^{(3)}$ achieves improvement in all experimental groups. This demonstrates that \sysname can adaptively optimize iterative results. 
We observe that with more iterations and increase the number of rewrites $g$, the attack effectiveness grow. Given this said, the incurred cost is slightly higher (see RQ4). As a practical setting, we recommend setting the number of iterations to 3, recovering 15\%$\sim$107\% more similarity with extra 1/4 remote quote and 3$\times$ local resource consumption.

\bheading{Query Accumulation.}
Additionally, as the query set $\mathbf{Q}$ continuously updates, subsequent recall results also improve. We retrieve the test data on the new $\mathbf{Q}^{(3)}$ again, achieving significant improvement in initial context $\hat{\mathbf{c}}^{(0)}$. This proves that as attacks progress, \sysname's probing of semantic space becomes increasingly comprehensive, further enhancing subsequent attack effectiveness.

\begin{table}[t]
\centering\small
\setlength{\arraycolsep}{3pt}
\renewcommand{\arraystretch}{0.6}
\caption{Improvement comparison across iterations.}
\resizebox{0.95\columnwidth}{!}{
\begin{tabular}{cccc||ccc}
\toprule
& \multicolumn{3}{c||}{\textbf{ $\|\tilde{\mathbf{c}}^{(3)}\|-\|\tilde{\mathbf{c}}^{(0)}\|$}} & \multicolumn{3}{c}{\textbf{$\|\hat{\mathbf{c}}^{(3)}\|-\|\hat{\mathbf{c}}^{(0)}\|$}} \\
\cmidrule(l){2-4} \cmidrule(l){5-7}
& \textbf{NED} & \textbf{ROUGE} & \textbf{COS} & \textbf{NED} & \textbf{ROUGE} & \textbf{COS} \\
\midrule
\cnum{1} & +15.2\% & +13.1\% & +10.8\% & +58.2\% & +37.1\% & +17.5\% \\
\cnum{2} & +25.8\% & +19.7\% & +17.7\% & +58.2\% & +34.6\% & +16.9\% \\
\cnum{3} & +52.5\% & +16.2\% & +13.6\% & +43.8\% & +25.6\% & +14.8\% \\
\cnum{4} & +46.4\% & +13.4\% & +11.6\% & +42.3\% & +26.1\% & +14.9\% \\
\cnum{5} & +117.4\% & +65.6\% & +30.7\% & +18.7\% & +19.2\% & +16.0\% \\
\cnum{6} & +107.0\% & +61.4\% & +29.8\% & +11.1\% & +9.4\% & +10.0\% \\
\bottomrule
\end{tabular}
}\vspace{-2mm}
\label{tab:improvement_comparison}
\end{table}

\subsection{RQ4: Attack Cost}

The resource consumption of \sysname primarily consists of remote LLM service calls and local GPU resource usage. We explain the costs at each stage.

\ulitalic{Initial probe budget.}
We spent roughly \$40 collecting initial data from real-world LLM vendors, split between DeepSeek-V3 and GPT-4o API calls. Although GPT-4o is more expensive, its responses contain  about 1/3 the tokens of DeepSeek-V3, reducing attack difficulty. Thus, we allocated 5K queries to GPT-4o, half of that used for DeepSeek. For the local Qwen3 model, we used 80 GPU hours to generate 5K queries.

\ulitalic{Attack quota for LLM.}
The main resource consumption during attack execution comes from local GPU usage. Recovering 1,200 victim conversations required approximately 150 GPU hours across a week, with extra time for prompt rewriting and new trace collection. Despite using a powerful reasoning model, this stage incurred only \$15 in API cost, showing that reasoning introduces minimal overhead.

\ulitalic{Iterative query budget.}
Each rewritten prompts and traces are collected to update the $\mathbf{Q}$. After three reinforcement rounds with early stopping, each Experiment added fewer than 700 queries, costing about \$5 in total—roughly 1/14 to 1/7 of the initial budget. Since these queries enhance future attacks and are reusable, iterative refinement remains cost-effective.

\begin{tcolorbox}[
    colback=gray!10,
    colframe=black,
    boxrule=1pt,
    arc=3pt,
    left=2pt,
    right=2pt,
    top=2pt,
    bottom=2pt
]
\textbf{Answer to RQs:} Using a \$55 budget, \sysname recovered 65\%/80\% of responses/prompts on medical data and 80\%/55\% on legal data across 1200 victim conversations, maintaining strong generalization on scaling and OOD data. Three iterations of refinement improved attack accuracy by up to 100\% while using only 1/4 of the remote quota, with all data being reusable for subsequent attacks.

\end{tcolorbox}

\section{Discussion}
\label{sec:discussion}

\bheading{Defense.}
To fully address the issue of defense, our analysis progresses sequentially: from the limitations of existing defenses, to the effectiveness of simulated advanced defenses, and concluding with future mitigations.

\ulitalic{Existing Defenses.}
Our primary goal was to assess the practical, real-world threat landscape by targeting defenses currently deployed by major vendors, which often prioritize cost and compatibility.
As we metioned in introduction, because  active defenses require modifying streaming protocols and Server Sent Event(SSE) interfaces, most
downstream applications cannot implement such measures.
Passive defenses is much easy to widely deployment, but excessive tokens per chunk significantly degrade
user experience, limiting it's defense strength.
We believe this focus is a key contribution, as it calibrates the actual risk users face today.

\ulitalic{Simulating Advanced Defenses.} 
To evaluate the advanced defense's effectiveness, we use topic recovery to show how many information will be exposed under dummy packet and padding. Firstly, we simulated dummy packets by inserting copies of existing packets at random positions in the traffic sequence. The result is shown in Fig.~\ref{tab:dummy_padding_impact}. This approach had a minimal impact, with recovery accuracy decreasing by an average of only 2\% when up to 20\% dummy packets were injected. Meanwhile, padding significantly degrades the effectiveness of Trace B, reducing the overall recovery performance to the same level as using Trace A alone in Sec.~\ref{sec:trace_eval}.

\ulitalic{Future Defenses.} 
The key limitation of advanced defenses is their deployment difficulty, which often requires tweaking protocols. Therefore, we expect to propose more systematic defenses, integrating prompt safety check~\cite{rebedea-etal-2023-nemo}, oblivious https~\cite{singanamalla2021oblivious}, \etc, into a toolkit like Apple PCC~\cite{ApplePCC}, to become a standard security component for LLM applications and serving high-security-requirement scenarios.

\begin{table}[t]
\centering\small
\setlength{\arraycolsep}{3pt}
\renewcommand{\arraystretch}{0.6}
\caption{Topic Recovery Simulation of the impact of dummy packets and padding. Values */* denote Top-1/Top-3 accuracy (\%).}
\resizebox{0.9\columnwidth}{!}{
\begin{tabular}{lcccccc}
\toprule
\textbf{Dummy rate} & \textbf{1\%} & \textbf{5\%} & \textbf{10\%} & \textbf{20\%} & \textbf{50\%} & \textbf{100\%} \\
\midrule
\cnum{2}-legal & 94.09/ & 93.99/ & 94.39/ & 93.88/ & 23.65/ & 12.23/ \\
Trace A+B & 99.18 & 98.88 & 99.29 & 99.18 & 48.83 & 32.01 \\
\midrule
\cnum{2}-med & 98.62/ & 98.34/ & 98.16/ & 97.98/ & 22.54/ & 14.63/ \\
Trace A+B & 99.82 & 99.45 & 99.45 & 99.36 & 54.46 & 38.64 \\
\midrule
\cnum{4}-legal & 90.60/ & 91.11/ & 92.48/ & 89.91/ & 90.09/ & 38.63/ \\
Trace A+B & 97.09 & 97.61 & 97.95 & 96.92 & 97.26 & 66.15 \\
\midrule
\cnum{4}-med & 90.50/ & 94.00/ & 94.50/ & 95.75/ & 82.50/ & 32.50/ \\
Trace A+B & 97.00 & 98.75 & 98.00 & 99.25 & 94.50 & 67.50 \\
\midrule
\cnum{5}-legal & 86.98/ & 90.93/ & 81.66/ & 77.12/ & 17.16/ & 13.02/ \\
Trace A+B & 93.49 & 95.27 & 92.90 & 92.50 & 40.04 & 33.14 \\
\midrule
\cnum{5}-med & 68.84/ & 93.95/ & 92.33/ & 30.00/ & 23.95/ & 31.86/ \\
Trace A+B & 91.86 & 98.60 & 97.91 & 63.02 & 50.47 & 66.51 \\
\bottomrule
\end{tabular}
}
\label{tab:dummy_padding_impact}
\end{table}

\bheading{Real-world Network Robustness.}
All our experiments were conducted on live, multi-tenant commercial services (\eg, OpenAI's API), not in a sanitized lab environment. The network noise, congestion, and jitter (which we observed in $>$5\% of connections) are already baked into our results, demonstrating \sysname's baseline robustness.

To further quantify this, we simulated  two types of network noise: (1) packet loss randomly removes a percentage of packets from the sequence, and (2) noise randomly swaps packet positions. Then we applied loss and noise from 0\% to 10\%—both separately and combined—to represent extreme network conditions. As shown in Table~\ref{tab:loss_noise_impact}, even under these disturbances, topic recovery using Trace A+B remained stable above 80\%, with only a minority of experimental groups showing lower accuracy.

\begin{table*}[t]
\centering\small
\setlength{\arraycolsep}{2pt}
\renewcommand{\arraystretch}{0.7}
\caption{Topic Recovery Simulation of the impact of packet loss and noise. Values */* denote Top-1/Top-3 accuracy (\%).}
\resizebox{2\columnwidth}{!}{
\begin{tabular}{c|ccccc|ccccc|ccccc}
\toprule
\multirow{2}{*}{\textbf{Loss/Noise Rate}} & \multicolumn{5}{c|}{\textbf{Loss \& Noise}} & \multicolumn{5}{c|}{\textbf{Only Loss}} & \multicolumn{5}{c}{\textbf{Only Noise }} \\
\cmidrule(lr){2-6} \cmidrule(lr){7-11} \cmidrule(l){12-16}
& 1\% & 3\% & 5\% & 7\% & 10\% & 1\% & 3\% & 5\% & 7\% & 10\% & 1\% & 3\% & 5\% & 7\% & 10\% \\
\midrule
\textbf{\cnum{1}-med, Trace A+B} & 98.07/ & 98.16/ & 97.88/ & 97.88/ & 97.61/ & 97.88/ & 98.25/ & 98.44/ & 98.25/ & 97.98/ & 98.07/ & 98.34/ & 98.62/ & 97.70/ & 97.06/ \\
& 99.26 & 99.36 & 99.17 & 99.54 & 99.45 & 99.36 & 99.26 & 99.63 & 99.63 & 99.45 & 99.36 & 99.54 & 99.72 & 99.36 & 99.17 \\
\midrule
\textbf{\cnum{1}-legal, Trace A+B}  & 95.21/ & 93.68/ & 93.68/ & 93.58/ & 89.91/ & 95.21/ & 94.09/ & 95.91/ & 94.60/ & 94.39/ & 94.80/ & 93.99/ & 93.37/ & 94.70/ & 89.09/ \\
& 99.39 & 98.98 & 99.18 & 99.18 & 98.67 & 99.08 & 99.29 & 99.39 & 99.18 & 98.98 & 99.08 & 99.08 & 98.88 & 99.08 & 98.06 \\
\midrule
\textbf{\cnum{4}-med, Trace A+B}  & 93.75/ & 93.25/ & 95.25/ & 85.75/ & 16.25/ & 94.00/ & 92.50/ & 94.50/ & 44.25/ & 70.25/ & 91.75/ & 93.25/ & 91.50/ & 94.50/ & 76.75/ \\
& 96.75 & 97.75 & 97.50 & 96.50 & 40.50 & 98.25 & 97.50 & 98.25 & 78.00 & 94.25 & 97.00 & 99.00 & 97.50 & 97.75 & 94.00 \\
\midrule
\textbf{\cnum{4}-legal, Trace A+B}  & 87.69/ & 90.60/ & 73.33/ & 79.32/ & 58.63/ & 90.94/ & 92.48/ & 92.14/ & 87.52/ & 89.91/ & 86.32/ & 90.60/ & 89.74/ & 91.79/ & 67.52/ \\
& 96.58 & 97.78 & 91.97 & 93.16 & 87.52 & 97.26 & 98.12 & 98.29 & 97.78 & 97.95 & 96.24 & 96.41 & 97.44 & 98.29 & 85.30 \\
\midrule
\textbf{\cnum{5}-med, Trace A+B}  & 91.52/ & 89.94/ & 89.35/ & 86.19/ & 55.82/ & 89.35/ & 90.93/ & 94.08/ & 91.32/ & 89.55/ & 89.15/ & 85.40/ & 88.76/ & 86.00/ & 85.40/ \\
& 96.65 & 95.07 & 95.07 & 94.67 & 84.22 & 95.27 & 95.07 & 96.65 & 95.07 & 95.27 & 94.87 & 93.69 & 93.49 & 95.07 & 93.89 \\
\midrule
\textbf{\cnum{5}-legal, Trace A+B}  & 91.52/ & 89.94/ & 89.35/ & 86.19/ & 55.82/ & 89.35/ & 90.93/ & 94.08/ & 91.32/ & 89.55/ & 89.15/ & 85.40/ & 88.76/ & 86.00/ & 85.40/ \\
& 96.65 & 95.07 & 95.07 & 94.67 & 84.22 & 95.27 & 95.07 & 96.65 & 95.07 & 95.27 & 94.87 & 93.69 & 93.49 & 95.07 & 93.89 \\
\bottomrule
\end{tabular}
}
\label{tab:loss_noise_impact}
\end{table*}

\section{Conclusion}
\label{sec:conclusion}


This paper systematically analyzes side-channel vulnerabilities in streaming responses of LLM applications. By identifying the limitations of existing defenses, we characterize six categories of real-world attack surfaces and quantify the leakage of multi-dimensional side channels. We also propose \sysname, a novel framework that recovers complete user-LLM conversations, maximizing the practical impact of this network-level leakage. Our findings highlight the urgent need for application-layer defenses, such as padding mechanisms, to be considered during streaming protocol design.

\normalem
\bibliographystyle{plain}
\bibliography{bib}

\begin{thebibliography}{10}

\bibitem{ApplePCC}
Private cloud compute: A new frontier for ai privacy in the cloud.
\newblock \url{https://security.apple.com/documentation/private-cloud-compute/}, 2024.
\newblock Accessed: 2025-10-29.

\bibitem{achiam2023gpt}
Josh Achiam, Steven Adler, Sandhini Agarwal, Lama Ahmad, Ilge Akkaya, Florencia~Leoni Aleman, Diogo Almeida, Janko Altenschmidt, Sam Altman, Shyamal Anadkat, et~al.
\newblock Gpt-4 technical report.
\newblock {\em arXiv preprint arXiv:2303.08774}, 2023.

\bibitem{harvey2025}
Harvey AI.
\newblock Harvey: Professional class ai for legal professionals.
\newblock \url{https://www.harvey.ai/}, 2025.

\bibitem{alhazbi2025llms}
Saeif Alhazbi, Ahmed Hussain, Gabriele Oligeri, and Panos Papadimitratos.
\newblock Llms have rhythm: Fingerprinting large language models using inter-token times and network traffic analysis.
\newblock {\em IEEE Open Journal of the Communications Society}, 2025.

\bibitem{carlini2024remote}
Nicholas Carlini and Milad Nasr.
\newblock Remote timing attacks on efficient language model inference.
\newblock {\em arXiv preprint arXiv:2410.17175}, 2024.

\bibitem{cloudflare2024mitigating}
Cloudflare.
\newblock Mitigating a token-length side-channel attack in our ai products.
\newblock \url{https://blog.cloudflare.com/ai-side-channel-attack-mitigated/}, 2024.

\bibitem{cloudflare2025workersai}
Cloudflare.
\newblock Demos and architectures - workers ai.
\newblock \url{https://developers.cloudflare.com/workers-ai/guides/demos-architectures/}, 2025.

\bibitem{microsoft2024azure}
Microsoft~Learn Community.
\newblock Azure gpt4o stream sends chunks at once in a short time with not good continuous flow.
\newblock \url{https://learn.microsoft.com/en-sg/answers/questions/1694655/azure-gpt4o-stream-sends-chunks-at-once-in-a-short}, 2024.

\bibitem{dyer2012peek}
Kevin~P Dyer, Scott~E Coull, Thomas Ristenpart, and Thomas Shrimpton.
\newblock Peek-a-boo, i still see you: Why efficient traffic analysis countermeasures fail.
\newblock In {\em 2012 IEEE symposium on security and privacy}, pages 332--346. IEEE, 2012.

\bibitem{fansi2022ddxplus}
Arsene Fansi~Tchango, Rishab Goel, Zhi Wen, Julien Martel, and Joumana Ghosn.
\newblock Ddxplus: A new dataset for automatic medical diagnosis.
\newblock {\em Advances in neural information processing systems}, 35:31306--31318, 2022.

\bibitem{siliconflow2025}
Silicon Flow.
\newblock Silicon flow: Ai model api service.
\newblock \url{https://siliconflow.com/}, 2025.

\bibitem{he2024foundation}
Yuting He, Fuxiang Huang, Xinrui Jiang, Yuxiang Nie, Minghao Wang, Jiguang Wang, and Hao Chen.
\newblock Foundation model for advancing healthcare: Challenges, opportunities and future directions.
\newblock {\em IEEE Reviews in Biomedical Engineering}, 2024.

\bibitem{glass2025health}
Glass Health.
\newblock Glass health: Ai diagnosis \& clinical decision support.
\newblock \url{https://glass.health/}, 2025.

\bibitem{henderson2023foundation}
Peter Henderson, Xuechen Li, Dan Jurafsky, Tatsunori Hashimoto, Mark~A Lemley, and Percy Liang.
\newblock Foundation models and fair use.
\newblock {\em Journal of Machine Learning Research}, 24(400):1--79, 2023.

\bibitem{hochreiter1997long}
Sepp Hochreiter and J{\"u}rgen Schmidhuber.
\newblock Long short-term memory.
\newblock {\em Neural computation}, 9(8):1735--1780, 1997.

\bibitem{kim2019make}
Jaehun Kim, Stjepan Picek, Annelie Heuser, Shivam Bhasin, and Alan Hanjalic.
\newblock Make some noise. unleashing the power of convolutional neural networks for profiled side-channel analysis.
\newblock {\em IACR Transactions on Cryptographic Hardware and Embedded Systems}, pages 148--179, 2019.

\bibitem{kwon2023efficient}
Woosuk Kwon, Zhuohan Li, Siyuan Zhuang, Ying Sheng, Lianmin Zheng, Cody~Hao Yu, Joseph~E. Gonzalez, Hao Zhang, and Ion Stoica.
\newblock Efficient memory management for large language model serving with pagedattention.
\newblock In {\em Proceedings of the ACM SIGOPS 29th Symposium on Operating Systems Principles}, 2023.

\bibitem{leviathan2023fast}
Yaniv Leviathan, Matan Kalman, and Yossi Matias.
\newblock Fast inference from transformers via speculative decoding.
\newblock In {\em International Conference on Machine Learning}, pages 19274--19286. PMLR, 2023.

\bibitem{lewis2020retrieval}
Patrick Lewis, Ethan Perez, Aleksandra Piktus, Fabio Petroni, Vladimir Karpukhin, Naman Goyal, Heinrich K{\"u}ttler, Mike Lewis, Wen-tau Yih, Tim Rockt{\"a}schel, et~al.
\newblock Retrieval-augmented generation for knowledge-intensive nlp tasks.
\newblock {\em Advances in neural information processing systems}, 33:9459--9474, 2020.

\bibitem{liu2024deepseek}
Aixin Liu, Bei Feng, Bing Xue, Bingxuan Wang, Bochao Wu, Chengda Lu, Chenggang Zhao, Chengqi Deng, Chenyu Zhang, Chong Ruan, et~al.
\newblock Deepseek-v3 technical report.
\newblock {\em arXiv preprint arXiv:2412.19437}, 2024.

\bibitem{lu2025fine}
Wei Lu, Rachel~K Luu, and Markus~J Buehler.
\newblock Fine-tuning large language models for domain adaptation: Exploration of training strategies, scaling, model merging and synergistic capabilities.
\newblock {\em npj Computational Materials}, 11(1):84, 2025.

\bibitem{nazari2024llm}
Najmeh Nazari, Furi Xiang, Chongzhou Fang, Hosein~Mohammadi Makrani, Aditya Puri, Kartik Patwari, Hossein Sayadi, Setareh Rafatirad, Chen-Nee Chuah, and Houman Homayoun.
\newblock Llm-fin: Large language models fingerprinting attack on edge devices.
\newblock In {\em 2024 25th International Symposium on Quality Electronic Design (ISQED)}, pages 1--6. IEEE, 2024.

\bibitem{oord2018representation}
Aaron van~den Oord, Yazhe Li, and Oriol Vinyals.
\newblock Representation learning with contrastive predictive coding.
\newblock {\em arXiv preprint arXiv:1807.03748}, 2018.

\bibitem{openai2023tiktoken}
OpenAI.
\newblock tiktoken: A fast bpe tokeniser for use with openai's models.
\newblock \url{https://github.com/openai/tiktoken}, 2023.
\newblock Accessed: 2024-01-01.

\bibitem{openai2024streaming}
OpenAI.
\newblock Streaming responses - openai api documentation.
\newblock \url{https://platform.openai.com/docs/guides/streaming-responses}, 2024.

\bibitem{lmsys2025arena}
LMSYS Org.
\newblock Chatbot arena: An open platform for evaluating llms by human preference.
\newblock \url{https://arena.lmsys.org/}, 2025.

\bibitem{rebedea-etal-2023-nemo}
Traian Rebedea, Razvan Dinu, Makesh~Narsimhan Sreedhar, Christopher Parisien, and Jonathan Cohen.
\newblock {N}e{M}o guardrails: A toolkit for controllable and safe {LLM} applications with programmable rails.
\newblock In Yansong Feng and Els Lefever, editors, {\em Proceedings of the 2023 Conference on Empirical Methods in Natural Language Processing: System Demonstrations}, pages 431--445, Singapore, December 2023. Association for Computational Linguistics.

\bibitem{shorten2019survey}
Connor Shorten and Taghi~M Khoshgoftaar.
\newblock A survey on image data augmentation for deep learning.
\newblock {\em Journal of big data}, 6(1):1--48, 2019.

\bibitem{singanamalla2021oblivious}
Sudheesh Singanamalla, Suphanat Chunhapanya, Jonathan Hoyland, Marek Vavru{\v{s}}a, Tanya Verma, Peter Wu, Marwan Fayed, Kurtis Heimerl, Nick Sullivan, and Christopher Wood.
\newblock Oblivious dns over https (odoh): A practical privacy enhancement to dns.
\newblock {\em Proceedings on Privacy Enhancing Technologies}, 4:575--592, 2021.

\bibitem{smith2025comprehensive}
Brandon Smith, Mohamed~Reda Bouadjenek, Tahsin~Alamgir Kheya, Phillip Dawson, and Sunil Aryal.
\newblock A comprehensive analysis of large language model outputs: Similarity, diversity, and bias.
\newblock {\em arXiv preprint arXiv:2505.09056}, 2025.

\bibitem{soleimani2025wiretapping}
Mahdi Soleimani, Grace Jia, In~Gim, Seung-seob Lee, and Anurag Khandelwal.
\newblock Wiretapping llms: Network side-channel attacks on interactive llm services.
\newblock {\em Cryptology ePrint Archive}, 2025.

\bibitem{song2024early}
Linke Song, Zixuan Pang, Wenhao Wang, Zihao Wang, XiaoFeng Wang, Hongbo Chen, Wei Song, Yier Jin, Dan Meng, and Rui Hou.
\newblock The early bird catches the leak: Unveiling timing side channels in llm serving systems.
\newblock {\em arXiv e-prints}, pages arXiv--2409, 2024.

\bibitem{thirunavukarasu2023large}
Arun~James Thirunavukarasu, Darren Shu~Jeng Ting, Kabilan Elangovan, Laura Gutierrez, Ting~Fang Tan, and Daniel Shu~Wei Ting.
\newblock Large language models in medicine.
\newblock {\em Nature medicine}, 29(8):1930--1940, 2023.

\bibitem{tuggener2020ledgar}
Don Tuggener, Pius Von~D{\"a}niken, Thomas Peetz, and Mark Cieliebak.
\newblock Ledgar: a large-scale multi-label corpus for text classification of legal provisions in contracts.
\newblock In {\em Proceedings of the twelfth language resources and evaluation conference}, pages 1235--1241, 2020.

\bibitem{chatbotui2024streaming}
Chatbot UI.
\newblock Streaming implementation discussion.
\newblock \url{https://github.com/mckaywrigley/chatbot-ui/issues/761}, 2024.

\bibitem{vercel2024playground}
Vercel.
\newblock Ai sdk playground.
\newblock \url{https://ai-sdk.dev/playground}, 2024.

\bibitem{vercel2025ai}
Vercel.
\newblock Vercel ai sdk: Ui package.
\newblock \url{https://github.com/vercel/ai/tree/main/packages/ai/src/ui}, 2025.

\bibitem{wang2024interactive}
Sheng Wang, Zihao Zhao, Xi~Ouyang, Tianming Liu, Qian Wang, and Dinggang Shen.
\newblock Interactive computer-aided diagnosis on medical image using large language models.
\newblock {\em Communications Engineering}, 3(1):133, 2024.

\bibitem{wang2020minilm}
Wenhui Wang, Furu Wei, Li~Dong, Hangbo Bao, Nan Yang, and Ming Zhou.
\newblock Minilm: Deep self-attention distillation for task-agnostic compression of pre-trained transformers.
\newblock {\em Advances in neural information processing systems}, 33:5776--5788, 2020.

\bibitem{wei2024privacy}
Jiankun Wei, Abdulrahman Abdulrazzag, Tianchen Zhang, Adel Muursepp, and Gururaj Saileshwar.
\newblock Privacy risks of speculative decoding in large language models.
\newblock {\em arXiv preprint arXiv:2411.01076}, 2024.

\bibitem{weiss2024your}
Roy Weiss, Daniel Ayzenshteyn, and Yisroel Mirsky.
\newblock What was your prompt? a remote keylogging attack on $\{$AI$\}$ assistants.
\newblock In {\em 33rd USENIX Security Symposium (USENIX Security 24)}, pages 3367--3384, 2024.

\bibitem{wolf-etal-2020-transformers}
Thomas Wolf, Lysandre Debut, Victor Sanh, Julien Chaumond, Clement Delangue, Anthony Moi, Pierric Cistac, Tim Rault, Rémi Louf, Morgan Funtowicz, Joe Davison, Sam Shleifer, Patrick von Platen, Clara Ma, Yacine Jernite, Julien Plu, Canwen Xu, Teven~Le Scao, Sylvain Gugger, Mariama Drame, Quentin Lhoest, and Alexander~M. Rush.
\newblock Transformers: State-of-the-art natural language processing.
\newblock In {\em Proceedings of the 2020 Conference on Empirical Methods in Natural Language Processing: System Demonstrations}, pages 38--45, Online, October 2020. Association for Computational Linguistics.

\bibitem{wu2025know}
Guanlong Wu, Zheng Zhang, Yao Zhang, Weili Wang, Jianyu Niu, Ye~Wu, and Yinqian Zhang.
\newblock I know what you asked: Prompt leakage via kv-cache sharing in multi-tenant llm serving.
\newblock In {\em Proceedings of the 2025 Network and Distributed System Security (NDSS) Symposium. San Diego, CA, USA}, 2025.

\bibitem{yang2024large}
Xiaoxian Yang, Zhifeng Wang, Qi~Wang, Ke~Wei, Kaiqi Zhang, and Jiangang Shi.
\newblock Large language models for automated q\&a involving legal documents: a survey on algorithms, frameworks and applications.
\newblock {\em International Journal of Web Information Systems}, 20(4):413--435, 2024.

\bibitem{yuan2022automated}
Yuanyuan Yuan, Qi~Pang, and Shuai Wang.
\newblock Automated side channel analysis of media software with manifold learning.
\newblock In {\em 31st USENIX Security Symposium (USENIX Security 22)}, pages 4419--4436, 2022.

\bibitem{zhang2024time}
Tianchen Zhang, Gururaj Saileshwar, and David Lie.
\newblock Time will tell: Timing side channels via output token count in large language models.
\newblock {\em arXiv preprint arXiv:2412.15431}, 2024.

\bibitem{zhao2023survey}
Wayne~Xin Zhao, Kun Zhou, Junyi Li, Tianyi Tang, Xiaolei Wang, Yupeng Hou, Yingqian Min, Beichen Zhang, Junjie Zhang, Zican Dong, et~al.
\newblock A survey of large language models.
\newblock {\em arXiv preprint arXiv:2303.18223}, 1(2), 2023.

\bibitem{zhao2024chatcad+}
Zihao Zhao, Sheng Wang, Jinchen Gu, Yitao Zhu, Lanzhuju Mei, Zixu Zhuang, Zhiming Cui, Qian Wang, and Dinggang Shen.
\newblock Chatcad+: Toward a universal and reliable interactive cad using llms.
\newblock {\em IEEE Transactions on Medical Imaging}, 43(11):3755--3766, 2024.

\bibitem{zheng2023judging}
Lianmin Zheng, Wei-Lin Chiang, Ying Sheng, Siyuan Zhuang, Zhanghao Wu, Yonghao Zhuang, Zi~Lin, Zhuohan Li, Dacheng Li, Eric.~P Xing, Hao Zhang, Joseph~E. Gonzalez, and Ion Stoica.
\newblock Judging llm-as-a-judge with mt-bench and chatbot arena, 2023.

\bibitem{zheng2024sglang}
Lianmin Zheng, Liangsheng Yin, Zhiqiang Xie, Chuyue~Livia Sun, Jeff Huang, Cody~Hao Yu, Shiyi Cao, Christos Kozyrakis, Ion Stoica, Joseph~E Gonzalez, et~al.
\newblock Sglang: Efficient execution of structured language model programs.
\newblock {\em Advances in neural information processing systems}, 37:62557--62583, 2024.

\bibitem{zheng2024inputsnatch}
Xinyao Zheng, Husheng Han, Shangyi Shi, Qiyan Fang, Zidong Du, Xing Hu, and Qi~Guo.
\newblock Inputsnatch: Stealing input in llm services via timing side-channel attacks.
\newblock {\em arXiv preprint arXiv:2411.18191}, 2024.

\end{thebibliography}



\appendix

\subsection{Inference Mechanism Speculation}\label{app:infer_mechanism_spec}

First, we hypothesize that both systems employ buffering mechanisms during streaming responses, which introduce certain noise artifacts. This is evidenced by occasional instances where the inference engine returns substantial amounts of data simultaneously while maintaining stable packet arrival intervals. This phenomenon likely results from optimization mechanisms such as parallel switching or dynamic batching during the inference process, which alter the otherwise stable time per output token. As illustrated in the Fig.~\ref{fig:token_distribution}, for our collected Deepseek and GPT data, some packets contain token counts that clearly exhibit outlier characteristics.
For GPT, since anomalous values consistently appear, the attack~\cite{weiss2024your} designed solely for pairing (2 tokens per packet) scenarios is ineffective.

Furthermore, repeated queries submitted in close temporal proximity clearly trigger caching mechanisms. For instance, when querying deepseek-v3, the second response is notably faster, likely due to prefill reuse leading to accelerated response generation. GPT-o4-mini, conversely, encapsulates more chunks within individual packets for transmission. This optimization also obscures the traces, making it challenging to infer conversational content while reuse the cache.

\ulitalic{Case Study.} 
Particularly, Deepseek's trace A exhibits consistent patterns. For example, packets containing 3 tokens demonstrate high content similarity, such as groups like (`$\backslash$n$\backslash$n',`1',`.') and (`$\backslash$n',`\ ',`*'). Therefore, the appearance of `3' in Deepseek's Trace A represents the initiation of new lines or separate content sections. This suggests specific token optimization mechanisms within Deepseek's architecture. No such phenomena were observed in GPT's behavior, indicating fundamental differences in their respective streaming strategies and token processing approaches.

\subsection{Analyzing Alternative Chunk Contents}
\label{app:variants}

In main text, we focus primarily on the standard chunk content format commonly used in production environments, which involves default streaming responses without additional parameters beyond \texttt{stream=true}. Per our observation, these streaming responses may have varying formats. Below, we categorize these variants into three types and explain that these variants still contain token-length information, often with even finer granularity.

\bheading{Additional Content Output.} We observed that under the same specifications, some third-party platform APIs return field content formats in streaming calls that differ from mainstream platforms like GPT. For instance, Silicon Flow enables the ``usage'' information in each chunk, which is null in mainstream APIs:

\begin{lstlisting}[basicstyle=\ttfamily\small, breaklines=true, breakatwhitespace=true, upquote=true]
usage: {
 prompt_tokens: 123,
 completion_tokens: 1,
 total_tokens: 124
}
\end{lstlisting}

We clarify that the token length information can be recovered by adversaries:
First, an adversary monitoring user sessions can estimate the character count of prompts from user request packets, thereby gaining further estimation of \texttt{prompt\_tokens}. Also, established connections return status confirmation packets, which also contain usage information. Moreover, the adversary has complete knowledge of the \texttt{completion\_tokens} field content in each packet, enabling precise inference of token length in most ranges. Errors only occur in a small interval after \texttt{total\_tokens} transitions from 999 to 1000.

\bheading{Additional Fields.} Engines like vLLM also add configuration fields such as \texttt{reasoning\_content}. These fields often do not change dynamically, thus they do not affect token length inference.

\bheading{Additional Request Parameters.} Some users may require more detailed output content, which can be achieved by using additional parameters in their requests. A typical example is the \texttt{log\_probability} parameter, which returns log probability information for each generated token in the API response, helping users debug outputs. In the returned packets, when token probability is extremely high, the \texttt{top\_logprobs} field content is as follows:

\begin{lstlisting}[basicstyle=\ttfamily\small, breaklines=true, breakatwhitespace=true,upquote=true]
logprobs:{content:
    [{token: other,
    logprob:0.0,
    bytes:[32,111,116,104,101,114],
    top_logprobs:[]}]}
\end{lstlisting}

Whereas under normal circumstances, it is as follows:

\begin{lstlisting}[basicstyle=\ttfamily\small, breaklines=true, breakatwhitespace=true, upquote=true]
logprobs:{content:
    [{token: two,
    logprob:-0.004620472434908152,
    bytes:[32,116,119,111]}]}
\end{lstlisting}

Thus, when \texttt{top\_logprobs} is enabled, the length difference between adjacent packets actually represents content with finer granularity than token length.
This fine-grained difference $\Delta = 2(\text{len}(t_1) - \text{len}(t_2)) + |\text{bytes}_1| - |\text{bytes}_2|$ provides richer side-channel information than simple token counting, where $|\text{bytes}_i| = \text{len}(\text{encode}(t_i))$ represents the byte length after token encoding. Due to significant differences in UTF-8 encoding lengths across different tokens, byte-length differences between adjacent response packets can reveal trace characteristics finer than token length.

\subsection{Dataset Organization}\label{app:dataset}

We use two domain-specific datasets for evaluating our attack framework: DDXPlus (medical) and LEDGAR (legal). At this step, both datasets are structured to create realistic LLM conversation scenarios with clear classification labels.

\bheading{DDXPlus Dataset.}
DDXPlus is a large-scale synthetic medical dataset containing approximately 1.3 million patients, designed for automatic symptom detection and diagnosis systems. The dataset includes differential diagnoses along with ground truth pathology, symptoms, and antecedents for each patient. Each patient's data is characterized by socio-demographic information, pathologies, and a comprehensive set of evidence (symptoms and antecedents) that can be binary, categorical, or multi-choice.

\bheading{LEDGAR Dataset.}
LEDGAR is a multi-label corpus containing legal provisions from contracts obtained from SEC filings, with over 12,000 labels annotated in almost 100,000 provisions across 60,000 contracts. For our classification task, we focus on a subset with 100 provision types following the LexGLUE benchmark.

\end{document}